\documentclass[11pt,a4paper]{article}
\usepackage{jheppub}

 \usepackage{tikz}
\usepackage{graphicx,verbatim}
\usepackage{amsmath}
\usepackage{amsfonts}
\usepackage{amssymb}
\usepackage{psfrag}
\usepackage{accents}

\usepackage[colorinlistoftodos]{todonotes}
\usepackage{epstopdf}

\def\Lie{\mathcal{L}}
\def\ps{{\Gamma}_{\rm cov}}
\def\psneh{{\Gamma}_{\rm cov}^{\rm NEH}}

\def\scri{\mathcal{I}}
\def\scrip{\scri^{+}}

\def\scripm{\scri^{\pm}}

\def\qo{\mathring{q}}

\def\qoub{\mathring{\underbar{q}}}

\def\vo{\mathring{v}}
\def\so{\mathring{s}}
\def\phio{\mathring{\phi}}
\def\chio{\mathring{\chi}}

\def\ello{\mathring{\ell}}

\def\no{\mathring{n}}
\def\gcirc{{}^{\circ}\!g}
\def\Do{\mathring{D}}

\def\epsilono{\mathring{\epsilon}}

\def\epsilonone{\epsilon^{\prime}}
\def\epsilontwo{\epsilon^{\prime\prime}}
\def\Ho{\mathring{\Delta}}

\def\hone{{}^{1}h}
\def\honedot{{}^1\dot{h}}
\def\ubhone{{}^{1}\underbar{h}} 

\def\htwo{{}^{2}h}
\def\ubhtwo{{}^{2}\underbar{h}}
\def\thetacirc{{}^\circ\theta_{(\ell)}}
\def\thetaone{\theta^\prime_{(\ell)}}
\def\thetatwo{\theta^{\prime\prime}_{(\ell)}}
\def\sigmacirc{{}^\circ\!\sigma^{(\ell)}}
\def\sigmaone{\sigma^{\prime\,(\ell)}}
\def\qcirc{{}^\circ\!q}
\def\epsiloncirc{{}^o\!\epsilon}

\def\psio{\mathring{\psi}}

\def\ThetaH{\underline{\Theta}{}}

\def\rmd{\mathrm{d}}

\def\lcirc{l}
\def\ncirc{n}
\def\ellb{\bar\ell}
\def\omegab{\bar\omega}

\def\H{\Delta}
\def\Hul{\underline{\Delta}}
\newcommand{\pb}[1]{\hbox{\lower0.5ex\hbox{${}_{\leftarrow}$}}\kern-1.9ex{#1}}

\def\G{\mathfrak{G}}
\def\g{\mathfrak{g}}

\def\B{\mathfrak{B}} 

\def\j{\mathfrak{J}}

\def\S{\mathcal{S}}
\def\C{\mathcal{C}}

\def\F{\mathcal{F}}
\def\N{\mathcal{N}}

\def\t{\tilde}

\def\b{\bar}

\def\={\hat{=}}
\def\f{\frac}

\def\be{\begin{equation}}
\def\ee{\end{equation}}
\def\ba{\begin{eqnarray}}
\def\ea{\end{eqnarray}}

\begin{document}

\title{Charges and Fluxes on (Perturbed) Non-expanding Horizons}
 
\author[a]{Abhay Ashtekar,}
\emailAdd{ashtekar.gravity@gmail.com}
\author[a]{Neev Khera}
\emailAdd{neevkhera@psu.edu}
\affiliation[a]{Institute for Gravitation and the Cosmos \& Physics Department, Penn State, University Park, PA 16802, U.S.A.}
\author[b]{Maciej Kolanowski}
\emailAdd{maciej.kolanowski@fuw.edu.pl}
\author[b]{and Jerzy Lewandowski}
 \emailAdd{jerzy.lewandowski@fuw.edu.pl}
\affiliation[b]{Institute of Theoretical Physics, Faculty of Physics, University of Warsaw, Pasteura 5, 02-093 Warsaw, Poland}

\abstract{
\noindent
In a companion paper  \cite{akkl1} we showed that the symmetry group $\G$ of non-expanding horizons (NEHs) is a 1-dimensional extension of the Bondi-Metzner-Sachs group $\B$ at $\scrip$. For each infinitesimal generator of $\G$, we now define a charge and a flux on NEHs \emph{as well as perturbed} NEHs. The procedure uses the covariant phase space framework in presence of internal null boundaries $\N$ along the lines of  \cite{cfp,cp2019,speziale1,ww2021,freidel1}. However, $\N$ is required to be an NEH or a perturbed NEH. Consequently, charges and fluxes associated with generators of $\G$ are free of physically unsatisfactory features that can arise if $\N$ is allowed to be a general null boundary. In particular, all fluxes vanish if $\N$ is an NEH, just as one would hope; and fluxes associated with symmetries representing `time-translations' are positive definite on perturbed NEHs. These results hold for zero as well as non-zero cosmological constant. In the asymptotically flat case, as noted in \cite{akkl1}, \emph{$\scripm$ are NEHs in the conformally completed space-time} but with an extra structure that reduces $\G$ to $\B$. The flux expressions at $\N$ 
reflect this synergy between NEHs and $\scrip$. In a forthcoming paper, this close relation between NEHs and $\scrip$ will be used to develop gravitational wave tomography, enabling one to deduce horizon dynamics directly from the waveforms at $\scrip$.

}
\maketitle

\section{Introduction}
\label{s1}

The purpose of this series of papers is to investigate geometry and physics of quasi-local horizons (QLHs) and relate them to structures available at $\scrip$. QLHs can be thought of as world tubes of marginally trapped 2-spheres (MTSs). Therefore, they can be located quasilocally  --\emph{quasi}locally, rather than locally, because one of the null normals of the MTSs is expansion-free on an entire $\mathbb{S}^2$, not just in the neighborhood of a point. However, in contrast to event horizons (EHs), the notion is far from being global; in particular, it makes no reference to future infinity $\scrip$. QLHs are also free from the strange teleological features that are characteristic of EHs. Since prior knowledge of space-time structure to the infinite future is not needed to locate them, QLHs have been extremely useful in the numerical evolution of black hole space-times depicting gravitational collapse, as well as mergers of compact bodies (see, e.g., \cite{akrev,boothrev,gjrev,schnetter_2006,jmmr1,jmmr2,Gupta_2018,pookkolb2020horizons,prasad2021tidal}). They are also better suited to understand physical processes since, unlike EHs that can grow in flat regions of space-time in anticipation of infall of matter and radiation \emph{in the distant future}, growth and dynamics of QLHs is governed by local physics in their immediate vicinity \cite{ak-dh}. 

QLHs can be divided into two broad classes: Non-expanding horizons (NEHs) that represent black hole and cosmological horizons in equilibrium, and dynamical horizons (DHs) that  describe growing (or, in the quantum theory, shrinking) black holes. Numerical simulations show that soon after the common apparent horizon forms in a binary black hole merger, the initially rapid evolution of the DH slows down and it is then well described by a perturbed NEH. As in the companion paper \cite{akkl1}, here we will restrict ourselves to NEHs and perturbed NEHs. DHs will be discussed in subsequent investigations.

Let us begin by summarizing the discussion and results of \cite{akkl1} that we will need on our analysis. We will only recall the salient features; precise definitions and proofs can be found in \cite{akkl1,afk,abl1}. An NEH $\H$ is a null, 3-dimensional sub-manifold --possibly a boundary-- of space-time, topologically $\mathbb{S}^2\times \mathbb{R}$, each of whose null normals is expansion-free. Thus, every 2-sphere cross-section $C$ of $\H$ is an MTS. Let $q_{ab} := g_{\pb{ab}}$ be the pull-back of the space-time metric to $\H$. Since $\H$ is null, $q_{ab}$ has signature $0, +,+$. The NEH $\H$ is naturally endowed with some additional geometric structures:\\
(i) One can always choose the null normals to $\H$ to be affinely parametrized geodesic vector fields $\lcirc^a$. Any two such normals are related by $\lcirc^{\prime\,a} = f\, \lcirc^a$ where $f$ satisfies $\Lie_{\lcirc} f =0$.\\
(ii) The space-time derivative operator $\nabla$ compatible with $g$ induces a unique derivative operator $D$ on $\H$ via pull-back. $D$ `interacts' with the degenerate metric $q_{ab}$ and the null normals $\lcirc^a$ in an interesting fashion. On $\H$ we have:
\be  D_a  q_{bc} = 0 \quad {\rm and} \quad D_a \lcirc^b = \omega_a\,\lcirc^b, \ee
for some $\omega_{a}$ which is called the `rotational 1-form' because it encodes the `angular momentum structure' of $\H$. ($\omega_a$ depends on the choice of the null normal $\lcirc^a$; but we will suppress this dependence for notational simplicity.) On $\H$, the pair $q_{ab},\, \omega_{a}$ satisfies
\be  q_{ab}\, \lcirc^b = 0,\,\,\,\Lie_{\lcirc} q_{ab} =0; \qquad {\rm and} \qquad \omega_a\, \lcirc^a =0, \,\,\, \Lie_{\lcirc} \omega_a =0\, . \ee
Thus, $q_{ab}$ and $\omega_a$ are pull-backs to $\Delta$ of covariant tensor fields on the 2-sphere $\Hul$ of integral curves of the null normals $\lcirc^a$. It is customary to go back and forth, regarding them as fields on $\H$ or on $\Hul$. 
Under $\lcirc^a \to \lcirc^{\prime\, a} = f \lcirc^a$ we have $\omega_a \to \omega^\prime_a = \omega_a + D_a \ln f$. \\
(iii) One can essentially exhaust the rescaling freedom in the choice of $\lcirc^a$ on $\H$ by requiring that $\omega_a$ be divergence-free on $\H$. This selects a small equivalence class of null normals, where two are equivalent if they are related by rescaling by a positive \emph{constant}. We will denote the equivalence class of these preferred geodesic null normals by $[\ellb^a]$ and the rotational 1-form they share by $\omegab_a$. Thus, $D^a \omegab_a =0$. As discussed in \cite{akkl1}, the \emph{NEH geometry} is encoded in the triplet $(q_{ab}, [\ellb^a], \omegab_a)$ that determines its intrinsic and extrinsic curvature which are neatly encoded in the real and imaginary part of the Weyl tensor component $\Psi_2$. Multipole moments provide an invariant characterization of these curvatures, and hence of the horizon geometry.\\
(iv) Every NEH $\H$ is equipped with a unique 3-parameter family of pairs $(\qo_{ab}, [\ello^a])$, consisting of \emph{unit, round} 2-sphere metrics $\qo_{ab}$ and equivalence classes of geodesic null normals $[\ello^a]$, where $\ello^{\prime\, a} \approx c \ello^a$ for some positive constant $c$. Every pair in this family is  
related to the pair $(q_{ab},\, [\ellb^a])$ on $\H$ via $\qo_{ab} = \psio^2 q_{ab}$ and $[\ello^a] = [\psio^{-1} \ellb^a]$ for some positive function $\psio$ on $\H$ satisfying $\Lie_\ell\, \psio =0$. The conformal factor $\psio$ relating the unit, round 2-sphere metrics $\qo_{ab}$ to the physical metric $q_{ab}$ (induced on $\H$ by $g_{ab}$) depends on  $q_{ab}$, and hence varies as we go from one NEH to another. However, the \emph{relative} conformal factors $\alpha$ \emph{relating one round metric to another} are universal because given a round metric $\qo_{ab}$ on an $\mathbb{S}^2$, there is precisely a 3-parameter family of round metrics that are conformally related to it. Thus, if $\qo_{ab} = \psio^2 q_{ab}$ and $\qo_{ab}^\prime = \psio^{\prime\,2}\, q_{ab}$, then $\qo_{ab}^\prime = (\psio^{\prime} \psio^{-1})^2 \,\qo_{ab}\, \equiv \, \alpha^2 \qo_{ab}$ and the conformal factor $\alpha$ relating the two round metrics is given by
\ba \label{alpha}
\alpha^{-1} &=& \alpha_0 + \sum_{i=1}^3 \, \alpha_i\, \hat{r}^i\, , \quad{\hbox{\rm for real constants $\alpha_0$ and $\alpha_i$,\,\,  with}} \nonumber\\
 \hat{r}^i &=& (\sin\theta\cos\phi,\, \sin\theta\sin\phi,\, \cos\theta)\qquad {\rm and}\qquad -\alpha_0^2 + \sum_{i=1}^3 (\alpha_i)^2 = -1\, , \ea   
where $(\theta, \phi)$ are spherical polar coordinates adapted to the first round metric $\qoub_{ab}$. In view of the constraints they satisfy, $\alpha_0$ and $\alpha_i$ can be thought of as components of  \emph{unit, time-like vectors} in Minkowski space. Consequently, on any NEH the space of these round metrics $\qo_{ab}$ can be thought of as the 3-dimensional, unit, space-like hyperboloid in Minkowski space. As is manifest in Eq. (\ref{alpha}), conformal factors $\alpha$ \emph{relating these round metrics to one another} are the same for all NEHs. \smallskip

Thus, every NEH $\H$ is equipped with a 3-parameter family of pairs $(\qo_{ab},\, [\ello^a])$ related by $(\qo_{ab}^\prime,\, [\ello^{\prime\,a}]) = (\alpha^2 \qo_{ab}, [\alpha^{-1} \ello^a])$, where $\alpha$ is given by Eq. (\ref{alpha}). This structure is universal: It is common to all NEHs since it does not refer to any fields that vary from one NEH to another.%
\footnote{$q_{ab}$ and $\omegab_a$ are `physical fields' that vary from one NEH to another and they determine the NEH multipole moments that characterize the geometry of individual NEHs. They are analogs of, say, $\Psi^\circ_2$ and $\Psi^\circ_1$ at $\scrip$ that carry information of mass and angular momentum at $\scrip$.} 
We can fix this family on an abstract 3-manifold $\Ho$, topologically $\mathbb{S}^2 \times \mathbb{R}$. The NEH symmetry group $\G$ is the subgroup of the Diffeomorphism group of $\Ho$ that preserves its universal structure, i.e., the collection of pairs $(\qo_{ab},\, [\ello^a])$. Given any space-time $(M, g_{ab})$ with a `concrete' NEH $\H$, there is a diffeomorphism from $\H$ to $\Ho$ that maps the pairs $(\qo_{ab},\, [\ello^a])$ on $\H$ to those we fixed on $\Ho$. This diffeomorphism is unique up to the action of $\G$ on $\Ho$; the situation is completely analogous to that at null infinity $\scrip$ \cite{aa-radiativemodes,as}.

Consider a 1-parameter family of diffeomorphisms $d(\lambda)$ that preserve the NEH universal structure, generated by a vector field $\xi^a$ on $\H$. It maps any given pair $(\qo_{ab},\, [\ello^a])$ to another pair $(\qo^\prime_{ab} = \alpha^2(\lambda),\,  [\ello^{\prime\,a}] = [\alpha^{-1}(\lambda) \ello^a] )$, with $\alpha(0) = 1$. Taking the derivative w.r.t. $\lambda$ at $\lambda=0$ and using Eq. (\ref{alpha}) we obtain the equations that define symmetry vector fields $\xi^a$ on $\H$:
\be \label{main} \Lie_\xi\, \qo_{ab} = 2\phio\, \qo_{ab}\qquad {\rm and} \qquad \Lie_\xi\, \ello^a = -\big(\phio + k\big)\,\ello^a \, .\ee
Here $k$ is a constant, and $\phio = {\rmd \alpha(\lambda)/\rmd \lambda}|_{\lambda=0}$ satisfies the linearized version of Eq. (\ref{alpha}). Thus $\phio$ is given by a general linear combination $\sum_m a_m Y_{1,m}$ of the first three spherical harmonics of $\qo_{ab}$. If one sets $k=0$, Eq. (\ref{main}) becomes identical to that defining the infinitesimal BMS vector fields on $\scrip$ (the null normals $\ello^a$  on NEH  being replaced by the null normals $\mathring{n}^a$ in a Bondi conformal frame at $\scrip$). The constant $k$ arises in (\ref{main}) because 
on NEHs a symmetry vector field that leave $\qo_{ab}$ invariant --i.e., corresponds to $\phi =0$-- can still rescale the null normals $\ello^a$ by a constant since NEHs are equipped only with \emph{equivalence classes} $[\ello^a]$ of null normals. This vector field is $\xi^a = k \vo\, \ello^a$ (where $\vo$ is an affine parameter of $\ello^a$) and represents the `dilation symmetry' that is absent in the BMS group. Thus the NEH symmetry group $\G$ is a 1-dimensional extension of the BMS group $\B$.  As we discussed in \cite{akkl1}, the extension is inevitable since, for example, the static Killing field in the Schwarzschild space-time appears as a dilation --rather than a supertranslation-- on its NEH.

Finally, note that the 3-parameter family of pairs $(\qo_{ab}, [\ello^a])$ is completely analogous to the  3-parameter family of \emph{`Bondi-conformal frames'} $(\qo_{ab}, \no^a)$ at $\scrip$ which are related by $(\qo_{ab}^\prime,\, [\no^{\prime\,a}]) = (\alpha^2 \qo_{ab}, [\alpha^{-1} \no^a])$ where $\alpha$ is again given by Eq.~(\ref{alpha})\, (see, e.g., \cite{aa-yau}). The round metrics $\qo_{ab}$ and the associated null normals $[\ello^a]$ were introduced in \cite{akkl1} to define multipole moments at $\H$; at $\scrip$ they serve to decompose waveforms and other physical fields into spherical harmonics which facilitates numerical simulations and data analysis. Also, they enable one to decompose the BMS vector fields at $\scrip$ into supertranslations, rotations and boosts, thereby providing further intuition for the associated charges and fluxes. We will see that  same is true at NEHs. Nonetheless, the pairs $(\qo_{ab}, [\ello^a])$ are not essential; one can work just with the physical metric $q_{ab}$ and the preferred equivalence class of preferred null geodesic normals $[\ellb^a]$ on $\H$. The symmetry vector fields $\xi^a$ of (\ref{main}) can be equivalently defined by $\Lie_{\xi} q_{ab} = 2\phi\, q_{ab};\,\, \Lie_{\xi} \ellb^a = -(\phi +k)\, \ellb^a$ where $k$ is a constant and $\phi$ satisfies $\Lie_{\ellb} \phi =0$. However, while the explicit expression of $\phio$ is simple and transparent,  $\phi$ are solutions to a more complicated equation on the 2-sphere with the metric $q_{ab}$ both at $\scrip$ and $\H$. 

In this paper we will discuss charges and fluxes associated with the NEH symmetry vector fields $\xi^a$ using a covariant phase space framework. In the context of space-times admitting null internal boundaries, the phase space framework has been developed in two different broad directions. In the first, the phase space was constructed from solutions that admit an internal boundary representing an NEH, possibly with a rotational symmetry (see, e.g. \cite{afk,abl2,akrev,boothrev,gjrev}). The focus was on the obtaining a generalization of the first law of horizon mechanics. This law was first established 
by Bardeen, Carter and Hawking \cite{bch} for Killing horizons in globally stationary (and axisymmetric) space-times. The goal was to extend it to space-times that admit only an NEH which is not necessarily a Killing horizon (e.g., as in the Kastor-Traschen solutions \cite{kt,kt1,kt2}), and can even admit radiation arbitrarily close to the NEH (as in the Robinson-Trautman solution \cite{pc}). The first law of horizon mechanics was shown to arise as a necessary and sufficient condition for the evolution along NEH symmetries to be Hamiltonian (see, e.g., \cite{afk,abl2,booth,acs,cns,bf}). As a byproduct the framework yielded charges representing the NEH angular momentum and energy. In these treatments every permissible variation preserves the NEH boundary conditions, whence all fluxes across the horizon vanish. The first law was thus obtained in the `passive setting' in which the variation in the energy, area and angular momentum refer to two nearby space-times in the phase space, both admitting NEHs but with slightly different energies, areas and angular momenta about a fixed symmetry axis. 

In the second and more recent set of investigations (see, e.g., \cite{cfp,cp2019,speziale1,ww2021,freidel1}), the covariant phase space again consists of solutions with an inner boundary $\N$. However, now $\N$ can be \emph{any} null surface, not necessarily an NEH. This framework is mathematically attractive because of its generality. Physically, it  has the pleasing feature that it allows fluxes of gravitational radiation across the boundary. On the other hand, because $\N$ is now allowed to be \emph{any} null surface, the symmetry group --e.g., $\G_{\rm CFP}$ of \cite{cfp}-- is \emph{much} larger than the BMS group $\B$ and, apart from the fact that is is also a semi-direct product, its detailed structure is also quite different from $\B$. More importantly, the framework allows a wide variety of situations in which charges and fluxes associated with symmetry generators are difficult to interpret physically. In particular, the charges and even fluxes can be non-zero \emph{in Minkowski space}. 
 
Thus, the first strategy seems too restrictive, and the second, too general. In this paper, we will place ourselves `in between' the two approaches. We will focus just on the sub-manifold\, $\psneh$ \,of the covariant phase space $\ps$, at each point of which the internal boundary is an NEH (rather than just a null 3-manifold). Therefore, our boundary symmetry group will be $\G$ which, as discussed in detail in \cite{akkl1}, is closely related to $\B$ in its structure. But we will also consider first and second order perturbations around these $g\in \psneh$ that do \emph{not} preserve the NEH structure. The phase space framework enables one to calculate (charges and) fluxes associated with the generators $\xi^a$ of $\G$ to these orders in perturbation theory. Thus, our results apply only to space-times that admit NEHs and perturbed NEHs. While this is a strong restriction, now fluxes can be interpreted physically: they vanish for the background, just as one would expect. On perturbed NEHs the flux expressions mirror those for perturbations of stationary space-times at $\scrip$. In particular, the flux of energy carried by perturbations is positive definite to second order. Also, since Minkowski space does not admit an NEH, it does not lie on $\psneh$ and the framework does not assign charges and fluxes to null surfaces in it. Finally, this strategy also bypasses the technical issue noted in \cite{cfp}, associated with the lack of smoothness of a general null boundary $\N$, due to geodesic crossing caused by the infalling radiation in the full, non-linear theory. As we indicated in \cite{akkl1}, the most promising approach to extend our results beyond perturbation theory would be to replace the perturbed NEHs that approximate slowly evolving DHs by full-fledged DHs.

This paper is organized as follows. In Sec.~\ref{s2.1} we specify our covariant phase space $\ps$ and its sub-manifold $\psneh$ for which the inner null boundary $\N$ is an NEH, and in Sec.~\ref{s2.2} we recall the main ingredients needed to define charges and fluxes at $\N$. In Sec.~\ref{s3} we discuss charges at NEHs and show that all fluxes vanish, just as one would physically expect. We also discuss properties of charges and show that they do not have any spurious features. In particular, because the 4-dimensional subgroup of NEH translations is augmented by a `dilation' symmetry, a priori one might be concerned that tensions might arise. Through examples we show that they do not. In Sec.~\ref{s4} we discuss charges and fluxes on perturbed NEHs and show that they too have physically expected properties. In Sec.~\ref{s5} we summarize the main findings and sketch some directions for future work that are being pursued. For convenience of the reader, in Appendix \ref{a1} we recall the covariant phase space framework in presence of an internal null boundaries used in the main text. Appendix \ref{a2} discusses the extension of the symmetry vector fields on $\N$ to its neighborhood that is necessary to evaluate the associated charges.  

Our conventions are the following. Throughout, the underlying space-time is 4-dimensional, the metric $g_{ab}$ has signature -,+,+,+, and its torsion-free derivative operator is denoted by $\nabla$.  Curvature tensors are defined via:  $2\nabla_{[a}\nabla_{b]} k_c = R_{abc}{}^d k_d$, $R_{ac} = R_{abc}{}^b$, and $R=g^{ab}R_{ab}$. Unless otherwise stated, all fields are assumed to be smooth for simplicity. \emph{If there is a possibility of ambiguity, we will use $\=$ to denote equality that holds only at the boundary $\N$.} Finally, null normals of the inner boundary $\N$ are assumed to be future directed.

\section{The Phase space framework}
\label{s2}

In the asymptotically flat case, a covariant phase space framework for general relativity has been available in the literature for quite some time \cite{cw,abr,wz}. However, that framework focused on charges and fluxes associated with asymptotic symmetries and did not consider space-times with internal boundaries. Recently, that framework was extended to include space-times with null internal boundaries $\N$ (see in particular, \cite{cfp,cp2019,speziale1,ww2021,freidel1}). As explained in Sec.~\ref{s1}, we will be primarily interested in the sub-manifold $\psneh$ of $\ps$ consisting of solutions $g_{ab}$ for which $\N$ is an NEH $\H$ (see Fig.~\ref{fig:triangle}), and perturbations $\delta g_{ab}$ (around metrics $g_{ab}$ in $\psneh$) that represent weak gravitational waves carrying non-zero fluxes across this $\H$. Thanks to this restriction, charges and fluxes will have physically expected properties. 

\subsection{Kinematic structures and the covariant phase space}
\label{s2.1} 

This sub-section is divided in three parts. In the first we introduce the covariant phase space $\ps$ and in the second, its sub-manifold $\psneh$ we will focus on. In the third part we discuss generic tangent vectors $\delta g$ at points $g\in \psneh$. As remarked in Sec.~\ref{s1}, there have been several discussions of the phase space framework in presence of \emph{general} null boundaries in the recent literature \cite{cfp,cp2019,speziale1,ww2021,freidel1}. Therefore, to provide a global perspective, we have organized the presentation so as to bring out the relation between those frameworks and ours, focusing on \cite{cfp} for concreteness.

\subsubsection{The phase space $\ps$}
\label{s2.1.1}

Fix a 4-manifold $M$ with a boundary $\N$ that is topologically $\mathbb{S}^2 \times \mathbb{R}$, equipped with a set $\S$ of oriented vector fields $\lcirc^a$ such that:\\
(i) the vector fields $\lcirc^a$ provide a ruling of $\N$, so that the space of their integral curves is diffeomorphic to $\mathbb{S}^2$. They are `future complete' in the sense that their affine parameters $v_o$ tends to $\infty$ in the direction in which they are oriented; and,\\
(ii) if $\lcirc^a$ and $\lcirc^{\prime\,a}$ are both in $\S$ then\, $\lcirc^{\prime\, a}\, \= \,f \lcirc^a$\, where $f$  satisfies $f >0$ and $\Lie_{\lcirc} f \=0$ on $\N$, but is otherwise arbitrary.\\
The set $\S$ does not refer to any space-time metric; as will be clear from our definition of $\ps$, it will serve as the kinematical structure shared by all $g\in \ps$. The subgroup $\G_\N$ of ${\rm Diff}(\N)$ preserving this structure is very large. Straightforward analysis shows that it is a semi-direct product of the group generated by `vertical' vector fields $\kappa (\theta,\varphi)\,v \lcirc^a + s(\theta,\varphi) \lcirc^a$ --which is thus `worth two functions on a 2-sphere'-- with ${\rm Diff\,} \mathbb{S}^2$, the diffeomorphism group of the space of `generators' $\lcirc^a$ of $\N$. This is the symmetry group of $\ps$ associated with the internal null boundary $\N$.

Our covariant phase space $\ps$ will consist of vacuum solutions $(M, g_{ab})$ of Einstein's equations (possibly with a cosmological constant) such that\\
(1) $\N$ is a null boundary of $M$, and every $\lcirc^a \in \S$ is an affinely parametrized geodesic null normal. Thus $\lcirc^a\, \nabla_a \lcirc^b\, \=\,0$;\\
(2) For any given $\lcirc^a$, the co-vector $l_{a} :\= g_{ab}\, \lcirc^b$ is the same for all $g\in \ps$.\\
(3) The expansion $\theta_{(\lcirc)}$ of each null normal vanishes in the distant future at least as fast as $1/v^{(1+\epsilon)}$ in the limit $v\to \infty$. If this condition is satisfied by one $\lcirc^a$, it is satisfied by all.

Condition (2) implies that $\N$ is equipped with pairs $(\lcirc^a,\, l_{a})$ where any two are related by a simple rescaling: \,$(\lcirc^{\prime a},\, l_{a}^\prime) \= (f \lcirc^a,\, f l_{a})$\, with\, $f >0$\, satisfying\, $\Lie_{\lcirc} f \=0$. This is a metric independent, kinematic structure on $\N$. It is the same as that used in \cite{cfp} but with $\kappa$ of \cite{cfp} now set to zero: As condition (1) requires, all our vector fields $\lcirc^a$ are not only null but also affinely parametrized geodesic vector fields on $\N$ for all $g$ in $\ps$.
Condition (2) might seem restrictive; however, as remarked in \cite{cfp}, every metric (satisfying (1)) can be mapped by a gauge diffeomorphism to a metric satisfying this condition. The third condition is new and introduced because we are only interested in situations such as gravitational collapse and mergers in which $\N$ becomes an NEH in the distant future. \smallskip

Recall that at $\scrip$ expansions in Bondi-Sachs or Newman-Unti coordinates serve to bring out the asymptotic form of the space-time metric. Let us introduce Newman-Unti type coordinates in a neighborhood of $\N$ and provide a more explicit form of the metrics $g \in \ps$. Fix a $\lcirc^a$ on $\N$, and an affine parameter $v$ of $\lcirc^a$. Introduce angular coordinates $x^A$ with $A=1,2$ on each $v\, \=\,{\rm const}$ 2-sphere such that $\Lie_{\lcirc} x^A\, \=\,\,0$ on $\N$. 
The coordinates $(v,\, x^A)$ on $\N$ make no reference to any space-time  metric $g$. However we will need a specific metric $g\in \ps$ to  extend them to a neighborhood in a systematic manner. Fix a\, $g\in \ps$,\, denote by $n^a$ the other future directed null normal to each $v={\rm const}$ cross-section of $\N$, normalized so that $g_{ab} \lcirc^a n^b =-1$, and consider past pointing null geodesics satisfying $n^a \nabla_a n^b =0$. Parallel transport $v, x^A$ along these geodesics and denote by $r$ the affine parameter of $- n^a$, with $r=0$ on $\N$. Then, we have a chart $v,r,x^A$ in a neighborhood of $\N$ (in which the geodesics tangential to $n^a$ do not intersect). In this chart, the metric $g\in \ps$ we began with has the form
\begin{equation} \label{metric}
    {g}_{ab} \rmd x^a \rmd x^b  = - r^2\, \gamma\, \rmd v^2 +2 \, \rmd v \rmd r + 2 r\,\beta_A\, \rmd v \rmd x^A + q_{AB}\, \rmd x^A \rmd x^B \, ,
\end{equation}
$\gamma, \beta_A, q_{AB}$  being smooth functions of our coordinates. (Note that we have six free functions, just as one would expect, given there is a 4-dimensional freedom in the choice of coordinates.) The vector field $n^a$ is given by $n^a \partial_a = -\partial_r$ in the neighborhood and by construction it is null and geodesic in the neighborhood. We extend $\lcirc^a$ to the neighborhood via $\lcirc^a\partial_a = \partial_v$; it is null and geodesic only on $\N$.%
\footnote{Note also that $l_{a}\, \hat{=}\, \nabla_a r $. Exactness of $l_{a}$ at $\N$ is used in \cite{cfp} to ensure certain uniqueness (see their Eq.(5.28)).} 
Since $r$ vanishes on $\N$, Eq. (\ref{metric}) implies  
\be {g}_{ab} \rmd x^a \rmd x^b \, \= \, 2 \, \rmd v \rmd r + q_{AB}\, \rmd x^A \rmd x^B \, .\ee
Thus, on the 3-dimensional boundary $\N$, the freely specifiable components of any  metric $g \in \ps$ are $q_{AB}$, just as one would expect from a characteristic initial value problem based on two intersecting null hypersurfaces  \cite{rendall}. Finally, because of our condition (3) on the metrics $g\in \ps$, $q^{AB} \partial_v q_{AB}$ vanishes at least as fast as \,$1/v^{1+\epsilon}$\, as\, $v\to \infty$. 

Consider now another metric $g^\prime \in \ps$. Then, keeping the $\lcirc^a$ and $v$ on $\N$ the same, we can repeat the procedure and obtain coordinates $(v^\prime, r^\prime, x^{\prime\, A})$ in a neighborhood of $\N$. The metric $g^\prime$ will have the same form as in (\ref{metric}) in these new coordinates which, by construction, agree with $(v, r, x^A)$ \emph{on} $\N$. Therefore, the diffeomorphism that sends the primed coordinates to unprimed is identity on $\N$ and can be regarded as gauge also from the phase space perspective. It sends $g^\prime$ to a metric that has the same form as in (\ref{metric}), \emph{now in unprimed coordinates}. Thus, we can fix the coordinates $(v, r, x^A)$ and consider all metrics of the form given in (\ref{metric}). Then, this collection contains a representative from each equivalence class of gauge related metrics in $\ps$. In this sense, there is no loss of generality in restricting ourselves to a fixed chart $(v, r, x^A)$ and solutions of Einstein's equations which have the form (\ref{metric}) in this chart. (Indeed, this strategy is often followed at $\scrip$, where one often fixes coordinates $(u, r, x^A)$ and restricts oneself to solutions which have the Newman-Unti (or Bondi-Sachs) asymptotic form near $\scrip$ in those coordinates.) While this description of $g\in \ps$ is often convenient, it is also awkward to use while discussing the action of  symmetries in $\G_\N$ on $\ps$ since the coordinates $(v, r, x^A)$ are tied to a specific pair $(\lcirc^a, v)$ and the pair changes under the action of $\G_\N$.  \smallskip

\subsubsection{The sub-manifold\, $\psneh$\, of \, $\ps$}
\label{s2.1.2}

\begin{figure}[]
  \begin{center}
  \hskip0.2cm
    \includegraphics[width=5.2in,height=2in,angle=0]{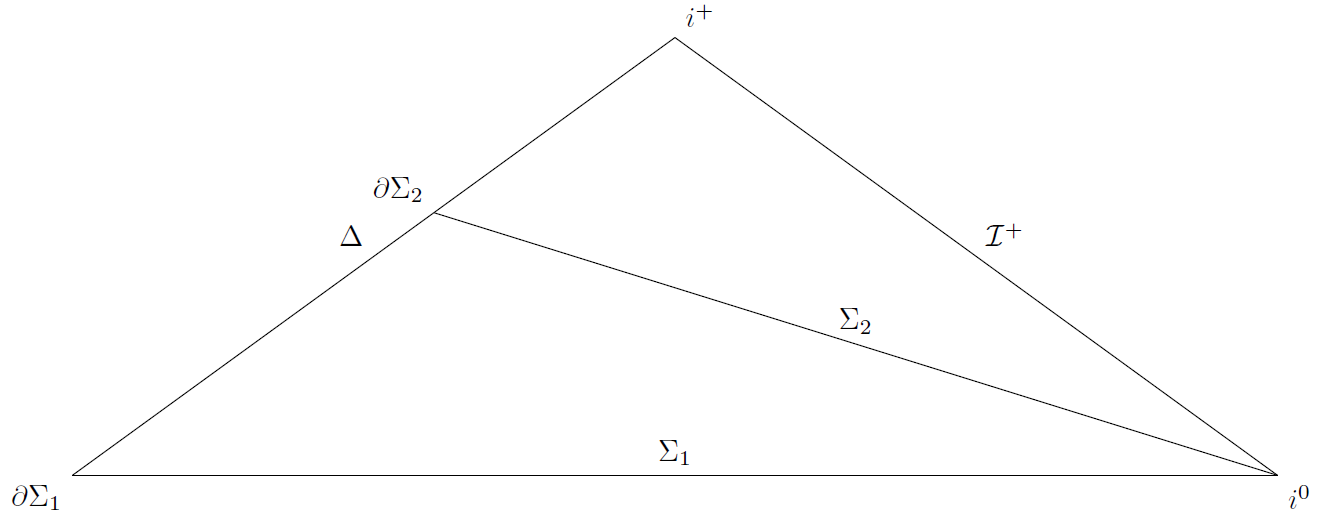}
    \caption{Part of the space-time that is relevant for the sub-manifold $\psneh$ of the full phase space $\ps$ that is of primary interest to our discussion. The inner boundary $\H$ is an NEH for each solution $g_{ab}$ in $\psneh$.\, $\Sigma_1$ and $\Sigma_2$ are partial Cauchy surfaces that intersect $\Delta$ in 2-spheres $\partial\Sigma_1$ and $\partial\Sigma_2$, respectively. Although this figure includes the asymptotic region including $\scrip$ and $i^o$, only a neighborhood of $\H$ is relevant for our detailed calculations.}     \label{fig:triangle}
    \end{center}
\end{figure}

As explained above, we wish to focus on a sub-manifold $\psneh$ of $\ps$ consisting on solutions $g$ for which $\N$ is an NEH (see Fig.~\ref{fig:triangle}). To specify this sub-manifold, we will again proceed in two steps, first fixing some additional kinematical structure on $\N$ and then specifying further requirements that $g\in \ps$ should satisfy in order to belong to $\psneh$.  

Recall from Sec.~\ref{s1} that each NEH comes equipped with a 3-parameter family of conformally related pairs $(\qo_{ab}, [\ello^a])$ consisting of unit, round 2-sphere metric $\qo_{ab}$ and an equivalence class of geodesic null normals $\ello^a$, where two of these normals are regarded as equivalent if they are related by rescaling with a positive constant $c$. Therefore, let us equip $\N$ with a collection $\C$ of such pairs, in addition to the set $\S$ we already fixed while defining $\ps$, such that\\
(i) Each $\ello^a$ belongs to the set $\S$; and \\
(ii) $(\qo_{ab}, [\ello^a]) \in \C$\, and\, $(\qo_{ab}^\prime, [\ello^{\prime\,a}]) \in \C$\, \,if and only if\, $(\qo_{ab}^\prime = \alpha^2\qo_{ab},\, [\ello^{\prime\,a}] = [\alpha^{-1} \ello^a])$ for a function $\alpha$ satisfying $\Lie_{\ello} \alpha =0$ and given by Eq. (\ref{alpha}). (Thus, there is precisely a 3-parameter family of functions $\alpha$.)

This is precisely the universal structure available on NEHs; in particular the functions $\alpha$ make no reference to any specific NEH geometry. As we showed in \cite{akkl1}, the family of diffeomorphisms that preserves $\C$ is the NEH symmetry group $\G$. This group is considerably smaller than the symmetry group associated with $\N$ in the full phase space $\ps$ which only has to preserve the entire set $S$ of vector fields $\lcirc^a$. This is because: (a) in addition to preserving the set $\S$ of vector fields $\lcirc^a$, the  diffeomorphisms in $\G$ have to preserve the much smaller subset of the vector fields $\ello^a$; (b) they also have to preserve the 3-parameter family of unit, round 2-sphere metrics $\{\qo_{ab}\}$; and, (c) they act on the conformally related pairs $(\qo_{ab}, [\ello^a])$ such that their action on $\qo_{ab}$ is appropriately coordinated with their action on  $[\ello^a]$.\smallskip

We can now specify the submanifold $\psneh$ of $\ps$. It will consist of solutions $g\in \ps$ that satisfy the following additional conditions:\\
(A) $\N$ is an NEH $\H$ with respect to the space-time metric $g_{ab}$;\\
(B) The pull-back $q_{ab}$ to $\N$ of $g_{ab}$ is conformally related to the unit, round 2-sphere metrics $\qo_{ab}$ in the collection $\C$:\, $\qo_{ab} = \psio^2 q_{ab}$ for some positive function on $\N$ satisfying $\Lie_{\ello}\, \psio =0$; and, \\
(C) The preferred equivalence class $[\ellb^a]$ of null geodesic normal to $\N$ selected by the NEH structure is related to the equivalence class $[\ello^a]$ in the pair $(\qo_{ab}, [\ello^a])$ by  $[\ello^a] = [\psio^{-1} \ellb^a]$.

Condition (A) implies that the expansion $\theta_{(\ello)}$ and shear $\sigma_{ab}^{(\ello)}$ of every null normal $\ello^a$ in $\C$ vanish on $\N$ (and therefore they vanish for all null normals to $\N$). Condition (B) on $g_{ab}$ requires that the metric $q_{ab}$ induced on $\N$ by $g_{ab}$ be conformally related to any of the round, unit 2-sphere metrics $\qo_{ab}$ in the collection $\C$. It then follows that all metrics $q_{ab}$ induced on $\N$ by $g\in \psneh$ are conformally related to one another. Similarly, Condition (C) requires that $g_{ab}$ be such that the preferred equivalence class $[\ellb^a]$ of normals on its NEH be related to the 3-parameter family of equivalence classes $[\ello^a]$ in $\C$ via the same conformal factor $\psio$ that relates the physical metric $q_{ab}$ to the unit round metrics $\qo_{ab}$. Therefore, at first glance, the restriction (B) and (C) on $g\in \psneh$ may appear to be  overly restrictive. However, further thought shows that they are not. Recall that the intrinsic geometry and angular momentum structure of any given NEH is captured in an invariant fashion by a set of multipole moments and the horizon geometry can be reconstructed from these moments. The `reconstruction argument' of  \cite{akkl1} implies that given \emph{any} set of (admissible) multipoles, there is a $g_{ab} \in \psneh$ with those multipoles. Hence from a physical perspective $\psneh$ is as rich as one would wish it to be. Together, conditions (A-C) imply that the sub-manifold $\psneh$ of $\ps$ consists of space-times $(M, g_{ab})$ in which $\N$ is an NEH $\H$, with the same universal structure as the one that was kinematically fixed through $\C$.

Finally, let us discuss how the requirement $g \in \psneh$ further constrains the form (\ref{metric}) of the metric $g_{ab}$ near $\N$.  Recall first that the NEH geometry is encoded in the triplet $(q_{ab}, [\ellb^a], \omega_a)$ where $q_{ab}$ is the intrinsic metric induced on $\H$ by the space-time metric $g_{ab}$ and $[\ellb^a]$ is the preferred equivalence class of null geodesic normals selected by the requirement that its rotational 1-form $\omegab_a$ --defined by ${\pb{\nabla_a}} \ellb\,{}^b = \omegab_a\, \ellb^a$-- is divergence free. Let us use an $\ellb^a \in [\ellb^a]$ to carry out the construction that led us to the form (\ref{metric}) of the metric. Then it follows that in the chart $(v, r, x^A)$ determined by that construction, $q_{AB}$ are the components of $q_{ab}$, and $-\f{1}{2}\beta_A$ of $\omega_a$. Therefore, $\beta_a$ is divergence-free w.r.t. $q_{ab}$. Finally, since $\Lie_{\ellb}\, q_{ab}\, \= \,0, \, \Lie_{\ellb}\, \omegab_a \,\=\,0$  and $q_{ab} \ellb^b\, \=0,\, \omegab_a \ellb^a\, \=\,0$ on any NEH $\H$ it follows that $q_{AB}$ and $\beta_A$ satisfy
\be \label{background}
    q_{AB,v} \, \=\, 0 \qquad {\rm and} \qquad  \beta_{A,v}\,\= \, 0\, . \ee

\subsubsection{Generic perturbations $\delta g$ around $g\in \psneh$}
\label{s2.1.3}

Let us now consider first order perturbations $\delta g_{ab}$ at points $g_{ab} \in \psneh$. These perturbations represent \emph{arbitrary} tangent vectors; they need not be tangential to $\psneh$. Therefore, generically, the NEH character of $\N$ is not preserved to first order. In particular, $\delta q_{AB}$ and $\delta \beta_A$ can have time-dependence. Nonetheless, since every $\lcirc^a \in \S$ that $\N$ is equipped with is a geodesic normal null to $\N$ for every $g\in \ps$, it follows in particular that these properties are preserved to the first order:
\be \label{gc1}
\delta g_{\pb{ab}}\, \lcirc^b\, \=\, 0 \qquad {\rm and} \qquad 
    \delta \left(\lcirc^a \nabla_a \, \lcirc^b\right)\, \=\, 0\, . 
\ee 
Next, for every $g\in \ps$ and $\lcirc^a \in \S$, the expansion $\theta_{(\lcirc)}$ falls-off at least as fast as $1/v^{1+\epsilon}$ for some $\epsilon >0$. The Raychaudhuri equation that holds for the geodesic null normals $\lcirc^a$ for any $g\in \ps$,
\be \label{Ray} \Lie_{\lcirc} \theta_{(\lcirc)} = - \textstyle{\f{1}{2}} \theta_{(\lcirc)}^2 - 
\sigma_{ab}^{(\lcirc)} \sigma_{cd}^{(\lcirc)} q^{ac} q^{bd}\, ,
\ee
then implies that the shear $\sigma_{ab}^{(\lcirc)}$ also falls-off at least as fast as $1/v^{1+\epsilon}$ for some $\epsilon >0$. Therefore for any linearized perturbation $\delta g$,\, $\delta \theta_{(\lcirc)}$ and $\delta \sigma_{ab}^{(\lcirc)}$ have the same fall-off as $v\to \infty$. Now, since for  $g\in \psneh$ the expansion $\theta_{(\lcirc)}$ as well as the shear $\sigma_{ab}^{(\lcirc)}$ determined by $g$ vanish everywhere on $\N$. Therefore, linearization of the Raychaudhuri equation about a $g \in \psneh$ implies $\Lie_{\lcirc} (\delta {\theta}_{(\lcirc)})\,\=\,0$. Since $\delta {\theta}_{(\lcirc)}$ vanishes in the limit $v \to \infty$, it follows that $\delta {\theta}_{(\lcirc)}\, \=\,0 $ everywhere on $\N$ for arbitrary perturbations $\delta g$ about any $g \in \psneh$. \emph{Thus, the perturbed expansion again vanishes to first order.} However, perturbed shear $\delta \sigma_{ab}^{(\lcirc)}$\emph{does not} vanish. Such perturbations represent weak gravitational waves. We will see in Sec.~\ref{s4} that the presence of such perturbations leads to non-trivial fluxes of physical quantities associated with NEH symmetries across $\N$, just as one would expect. Finally, even for perturbations $\delta g$ that are tangential to $\psneh$, the angular derivatives of $\delta q_{ab}$ and $\delta \beta_a$ are generically unconstrained. Such perturbations map the NEH geometry of $g$ to a nearby NEH geometry, with different multipoles. For example if $g$ is the Schwarzschild metric, then $\delta g$ could be a linearized Kerr metric on the background $g$. Under such perturbations, charges associated with the NEH symmetries change but there will be no fluxes, again just as one would expect. \smallskip

To summarize, the universal structure defined intrinsically on $\N$ consists of the set $\S$ of vector fields\, $\lcirc^a$\, specified in Sec.~\ref{s2.1.1}. The symmetry group preserving this structure $\G_\N$ has a semi-direct product structure like the BMs group $\B$, but the normal subgroup (of `vertical diffeomorphisms') is `worth two functions on a 2-sphere' rather than one, and the quotient is the infinite dimensional ${\rm Diff}\, \mathbb{S}^2$, rather than the six dimensional Lorentz subgroup. The full phase space $\ps$ consists of those solutions $g_{ab}$ of Einstein's equations for which $\N$ is a null 3-boundary of space-time, each $\lcirc^a\in \S$ serving as a geodesic null normal, and satisfying the condition that for any given $\lcirc^a$, the covariant normal $l_{b} := g_{ab}\, \lcirc^a$ is the same, independent of $g_{ab}$.%
\footnote{Physical considerations also led us to require that $\N$ becomes an NEH in the distant future for all $g\in \ps$. To bring out the similarity between structures on $\N$ and at $\scrip$, it is convenient to assume in addition that $g_{ab}$ is asymptotically flat in the $\Lambda =0$ case, although this restriction plays no direct role in our analysis.}
One can choose coordinates $(v, r, x^A)$ in a neighborhood of $\N$ such that every $g\in \ps$ is gauge related to the metric of the form (\ref{metric}). To define the sub-manifold $\psneh$ of $\ps$ that is of primary interest in our analysis, we introduced additional kinematic structure on $\N$: A collection $\C$ consisting of pairs $(\qo_{ab}, [\lcirc^a])$, specified in Sec.~\ref{s2.1.2}. The group preserving the full kinematic structure is $\G$, whose structure is discussed in detail in \cite{akkl1}. It is a 1-dimensional extension of the BMS group $\B$ and thus considerably smaller than $\G_\N$. The sub-manifold $\psneh$ of $\ps$ consists of those $g\in \ps$ for which $\N$ is an NEH such that the intrinsic metric $q_{ab}$ and the canonical equivalence class $[\ellb^a]$ of null normals on this NEH are conformally related to the pre-specified pairs $(\qo_{ab}, [\ello^a])$ in the manner spelled out in Sec.~\ref{s2.1.2}. Because of these additional conditions on $g_{ab}$ the metric coefficients in (\ref{metric}) are now constrained by Eq.  (\ref{background}). Finally we discussed the structure of generic perturbations $\delta g$ at points $g\in \psneh$. Our conditions on $\psneh$ imply that, to first order in perturbations, vector fields $\lcirc^a$ continue to be geodesic null normals to $\N$ as in Eq. (\ref{gc1}),\, and the expansion continues to vanish; $\delta \theta_{\lcirc} \=0$. However the first order shear $\delta \sigma_{ab}^{(\lcirc)}$ is generically non-zero on $\N$, just as one would expect of perturbations representing gravitational waves that traverse $\N$.
      
\subsection{New ingredients}
\label{s2.2}

To define charges and fluxes at internal, null boundaries, two new ingredients have to be added to the standard Hamiltonian framework. To spell them out, let us begin with the standard covariant phase space of general relativity \cite{cw,abr}. It consists of suitable solutions $g_{ab}$ of Einstein's equations equations on a given 4-manifold $M$, such that $(M, g_{ab})$ is globally hyperbolic. Given a solution $g$ in the covariant phase space and a tangent vector $\delta g$ at this point, the Lagrangian formulation provides us with a 3-form $\Theta_{abc}(g; \delta g)$ \emph{on space-time},
\be \label{Thetao}  
\Theta_{abc}\,(g;\delta g) = \frac{1}{16\pi G}\,\,\epsilon_{abc}{}^d\,(g^{ef}\nabla_d\,\delta g_{ef}-\nabla^e\delta g_{ed})\, , 
\ee
that depends linearly on $\delta g$. For globally hyperbolic space-times under consideration, the integral of $\Theta_{abc}(g;\delta g)$ over a Cauchy surface provides a 1-form  $\mathbf{\Theta} (g)$ \emph{on the covariant phase space}, whose exterior derivative yields the pre-symplectic 2-form $\mathbf{\Omega}(g)$. Since $\mathbf{\Theta} (g)$ is a symplectic potential on the phase space, the space-time 3-form $\Theta_{abc}(g;\delta g)$ will be referred to as the `obvious 3-form potential'. 

Now, given a pre-symplectic potential $\mathbf{\Theta}$, we can add to it the gradient of a phase space function to obtain another, equally viable pre-symplectic potential. This freedom translates to that of adding to $\Theta_{abc} (g; \delta g)$ a term of the form $\delta F_{abc}(g)$, where $F_{abc}(g)$ is a space-time 3-form that depends on the given metric $g$, and $\delta F_{abc}$ its directional derivative along $\delta g$ in the argument of $\Theta_{abc}$. 

The definition of charges and fluxes associated with the internal null boundary $\N$ requires us to eliminate this `gauge freedom' \cite{wz,cfp}. More precisely, one needs the pull-back $\ThetaH{}_{abc} (g; \delta g)$ to $\N$ of a `preferred 3-form potential'. This 3-form on $\N$ is selected by imposing a set of natural conditions, summarized in Appendix \ref{a1}. This is the first ingredient that one needs to define charges and fluxes. This `preferred' $\ThetaH{}_{abc} (g; \delta g)$ on $\N$ is given by  
\be \ThetaH_{abc} (g;\, \delta g) \, =\, {\Theta}_{\pb{abc}} (g; \delta g) - \f{1}{8\pi G}\,  \delta \big( (\theta_{(\lcirc)}\, \epsilon_{abc})(g)\big)\, . \ee 
Here ${\Theta}_{\pb{abc}} (g; \delta g)$ is the pull-back to $\N$ of $\Theta_{abc}\,(g;\delta g)$ of Eq. (\ref{Thetao}); $\theta_{(\lcirc)}$ denotes the expansion of the null normal $\lcirc$ to $\N$,\, as before; and $\epsilon_{abc}$, the volume 3-form on $\N$ defined by 
\be \epsilon_{abc} := n^d \epsilon_{dabc} \quad \hbox{\rm where $n^a$ is any vector field on $\N$ s.t. $\lcirc^a n^b\, g_{ab} =-1$} \, .\ee
Note that the term $(\theta_{(\lcirc)}\, \epsilon_{abc})$ is independent of the choice of the geodesic null normal $\lcirc^a$ and the associated $n^a$ used in its evaluation. Consequently, $\ThetaH_{abc}(\delta g)$ is also insensitive to this choice.

The second ingredient one needs to define charges and fluxes is a suitable extension $X^a$ of the symmetry vector fields $\xi^a$ on $\N$ to its neighborhood in the space-time manifold $M$. The actual calculations require only the first outward derivatives of the extension and a prescription to fix this derivative was given in \cite{cfp}. Since in our analysis $\N$ is restricted to be an unperturbed or a perturbed NEH $\H$, we can do more: we can provide an explicit form of the extension $X^a$ to the desired order by fixing a chart $(v,r,x^A)$ as in Sec.~\ref{s2.1.1}:
\be\label{X1} X = (v f_1 + f_2 ) \partial_v + H^A \partial_A - r f_1 \partial_r - r \t{X}^v\partial_v + r \tilde{X}^A \partial_A + r^2 \tilde{\tilde{X}}^r \partial_r \, . \ee
Here $f_1, f_2, H^A$ are functions only of the angular coordinates $x^A$. As one can see by setting $r=0$, the symmetry vector field $\xi^a$ on $\N$ is given by $\xi\,  \=\, (v f_1 + f_2 ) \partial_v + H^A \partial_A$. Thus, $f_1, f_2, H^A$ are completely determined by the given $\xi^a$. $\tilde{X}^v,\,\tilde{X}^A,\, \tilde{\tilde{X}}^r$ are new smooth functions (of $(v,r, x^A)$) that represent the freedom in the extension. However, as we will see in Secs.~\ref{s3} and \ref{s4}, the charges and fluxes associated with $\xi^a$ are insensitive to this freedom. The Lie algebra of these (equivalence classes) of extensions $X^a$ is the same as the Lie algebra $\g$ of our symmetry vector fields $\xi^a$ on $\H$. (For details on this extension and its properties, see Appendix~\ref{a2}.) Finally, the form of the extension given in (\ref{X1}) holds only in a neighborhood of $\N$. If one is interested in charges also at infinity, one would have assume that $X^a$ can be further extended so that it is an asymptotic symmetry. If one is interested in charges and fluxes only on $\N$ it is convenient to assume that the extended vector field $X^a$ vanishes outside a spatially finite region.  

With these two ingredients at hand, the expression of charges at the null boundary $\N$ are given by  \cite{cfp}:
\be \label{cfpcharge}
Q_\xi [C]\, =\,  Q^{\rm N}_\xi\, [C] - \f{1}{8\pi G}\, \oint_C \theta_{(\ell)}\, \xi^a \, \epsilon_{abc} \, \equiv 
- \f{1}{16\pi G}\,\Big( \oint_C \epsilon_{abcd} \nabla^c X^d + 2\, \theta_{(\ell)}\, \xi^c \, \epsilon_{cab}\Big)\, . \ee
The first term, $Q^{\rm N}_\xi\, [C]$, is the Noether charge that results if one uses the pre-symplectic potential $\Theta_{\pb{abc}} (g; \delta g)$, while the second term arises from the `gradient term' $\f{1}{8\pi G}\,  \delta \big( (\theta_{(\ell)}\, \epsilon_{abc})(g)\big)$ that was added to obtain the canonical $\ThetaH_{abc} (g; \delta g)$.  The extension $X^a$ of the symmetry vector field $\xi^a$ on $\N$ is needed to evaluate the Noether charge since its integrand requires the knowledge of the outward (i.e., radial) derivative of $X^a$ at points on $C$. We will find that only the part $-r\,f_1(x^A)\,\partial_r$ of the extension is needed in this calculation and $f_1(x^A)$ is determined directly by the symmetry vector field $\xi^a$ on $\N$. Therefore, the charge $Q_\xi [C]$ depends only on the space-time metric $g_{ab}$ and the symmetry vector field $\xi^a$; it is independent of additional structures we will introduce in Sec. \ref{s3} to make the physical interpretation of charges transparent.  Finally note that, if the symmetry vector field $\xi^a$ is either tangential to the cross-section $C$, or if $\N$ is an NEH for the metric $g$ under consideration, the second term vanishes and $Q_\xi [C]$ is given just by the Noether charge.

The flux across the portion $\N_{1,2}$ of $\N$ bounded by two cross-sections $C_1$ and $C_2$ (with $C_1$ to the future of $C_2$) is given, of course, by the integral of the exterior derivative of the 2-form integrand of $Q_\xi [C]$ over $\N_{1,2}$. One can show that this expression can be recast in the form 
\be \label{flux} \F_\xi\, [\N_{1,2}] =   \int_{\N_{1,2}} \!\! \ThetaH_{abc}\, (g; \delta_X g) \ee
that is sometimes more convenient to use. (Our conventions are such that $\F_\xi\, [\N_{1,2}]$ is the flux leaving space-time portion depicted in Fig.~\ref{fig:triangle} and falling inward.) All these considerations hold on the full phase space $\ps$. 

In Sec.~\ref{s3} we will evaluate these charges and fluxes for various symmetry generators by restricting ourselves to solutions $g$ for which $\N$ is an NEH and, in Sec.~\ref{s4}, for metrics $g$ for which it is a perturbed NEH.

\section{Charges and fluxes on NEHs}
\label{s3}

This section is divided into three parts. In the first we collect expressions that will be used in the subsequent subsections as well as in Sec.~\ref{s4}; in the second we calculate charges and fluxes on NEHs; and in the third we use explicit examples to illustrate that they are viable from physical considerations, and have interesting properties.

\subsection{General considerations} 
\label{s3.1}

To begin with, let us consider general metrics $g\in \ps$ for which $\N$ is \emph{not necessarily an NEH.} Fix \emph{any} geodesic null normal $\lcirc^a$, and a null vector field $\ncirc^a$ on $\N$ that is normal to the cross-section $C$ of interest and satisfies $\lcirc^a \ncirc^b g_{ab}\, \= \, -1$. Then, the expression of the Noether charge $Q^{\rm N}_\xi$ becomes:  
\ba \label{QN2} Q^{\rm N}_\xi\, [C] & \equiv& - \f{1}{16\pi G}\,\oint_C \epsilon_{ab}{}^{cd} \nabla_c X_d\, =\, - \f{1}{8\pi G}\,\oint_C \epsilon_{ab}\,\, \lcirc^c \ncirc^d\, \nabla_{[c} X_{d]} \nonumber\\
&=& - \f{1}{16\pi G}\,\oint_C \epsilon_{ab}\,\, \ncirc^d \big( \Lie_{\lcirc} X_d - \nabla_d (\lcirc \cdot X) \big)
\ea
where $\epsilon_{ab}$ is the area 2-form on $C$,\, $X^a$ is the extension (\ref{X1}) of the symmetry vector field $\xi^a$. To simplify this expression further, let us 
fix an affine parameter $v$ of this  $\lcirc^a$, and introduce the `Newman-Unti'-type chart $(v,r,x^A)$ in a neighborhood of $\N$ as in Sec.~\ref{s2.1}. Then, we have $\lcirc^a\partial_a = \partial_v$ and $\ncirc^a\partial_a = -\partial_r$ in a neighborhood of $\N$, so that $\Lie_{\lcirc} \ncirc^a =0$. Using this fact in (\ref{QN2}) and the form (\ref{metric}) of the space-time metric in these coordinates, we have:
\ba \label{QN3} Q^{\rm N}_\xi\, [C] &=& - \f{1}{16\pi G}\,\oint_C \epsilon_{ab}\,
\big[ \Lie_{\lcirc} (g_{cd} X^c\, \ncirc^d)\, -\, \Lie_{\ncirc} (g_{cd} X^c\, \lcirc^d)\big]\nonumber\\
&=& \f{1}{8\pi G}\,\oint_C \epsilon_{ab} \big(f_1 \, - {\textstyle\f{1}{2}}\, \beta_a\,H^a \big)\ea
where ($\epsilon_{ab}, \, \beta_a$)  are determined by the metric $g_{ab}$ while $f_1,\, H^a$ are determined by symmetry vector field $\xi^a$ on $\N$ itself (see the expression (\ref{X1}) of $X$). The first expression in (\ref{QN3}) makes it manifest that the charge $Q^{\rm N}_\xi\, [C]$ depends only on the space-time metric and the extension $X^a$ of the symmetry vector field while the second expression makes it clear that all the necessary information in the extension $X^a$ is available already in $\xi^a$ on $\N$; auxiliary information such as the fields that remain undetermined in the extension (\ref{X1}) do not enter. With this expression of $Q^{\rm N}_\xi\, [C]$ at hand, the total charge $Q_{\xi}[C]$ of Eq.~(\ref{cfpcharge}) becomes:
\be \label{Q}
Q_\xi\, [C] = \f{1}{8\pi G}\,\oint_C  \big(f_1\, -\, \textstyle{\f{1}{2}}\beta_a\,H^a \big)\epsilon_{ab}\,  -\, \,\theta_{(\lcirc)}\, \xi^c\, \epsilon_{cab} \, . \ee
Eq.~(\ref{Q}) holds for \emph{any} choice of the geodesic null normal $\lcirc^a$ (in the universal structure at $\N$ spelled out in Sec.~\ref{s2.1.1}). Since $(\theta_{(\lcirc)}\,\epsilon_{cab})$ is independent of the choice of $\lcirc^a$, the charges depend only on the infinitesimal symmetry generator $\xi^a$ and \emph{physical} fields $\epsilon_{ab}$,\, $\beta_a$ and $(\theta_{(\lcirc)}\, \epsilon_{cab}$),  induced by the space-time metric at $C$.

To associate physical interpretation to charges and corresponding fluxes it is convenient to decompose the vector fields $\xi^a$ into various parts, just as one does at null infinity $\scrip$. As discussed in \cite{akkl1}, $\xi^a$ can be written as a sum of a dilation $d^a$, a supertranslation $S^a$ a rotation $R^a$ and a boost $B^a$. However, this decomposition is not defined invariantly: It requires us to choose a round metric $\qo_{ab}$ from the universal structure of Sec.~\ref{s2.1.2}. We also need to choose a vector field $\ello^a$ in the equivalence class $[\ello^a]$ corresponding to $\qo_{ab}$, and an affine parameter $\vo$ of $\ello^a$ to display various pieces of the decomposition explicitly.
\footnote{\label{fn5} As we pointed out in sec. \ref{s1}, the situation is completely analogous to that at $\scrip$. In our case, one can select a unique $(\qo_{ab}, [\ello^a])$ on $\H$ by requiring that the `area dipole moment' vanish --i.e., $\oint_C \mathring{Y}_{1m} \epsilon_{ab} =0$ on $\H$ \cite{akkl1}. Given a space-time with an NEH, one can use this $\qo_{ab}$. However, we will not confine ourselves to this choice because this $\qo_{ab}$ varies from one space-time to another.}
 But as we will see, these two additional choices play a minor role. Given $(\qo_{ab}, \ello^a, \vo)$, we can write a general element $\xi^a$ of the symmetry Lie algebra $\g$ as 
\ba
\label{xi2} \xi^a &=& k \,\vo\,\ello^a  + \so\, \ello^a \,+\, \epsilono^{ab} \Do_b \chio\, + \, (\phio\, \vo\, \ello^a +\qo^{ab} \Do_b \phio)\, \,  \nonumber\\ 
&\equiv& d^a + S^a + R^a + B^a \ea
Thus the dilation $d^a$ is labelled by a constant $k$,\, the supertranslation $S^A$, by a function $\so(x^A)$ of angles, the rotation $R^a$ by a function $\chio (x^A)$ and the boost $B^a$ by a function $\phio(x^A)$, where $\chio(x^A)$ and $\phio(x^A)$ are both linear combinations of the $\ell =1$ spherical harmonics defined by the $\qo_{ab}$ we fixed. (For details, see Sec.~3 of \cite{akkl1}.) Note that the combination $\vo\, \ello^a$ is unchanged under rescalings of $\ello^a$ and $\so$ is a function with conformal weight $1$. Consequently, if we use a different conformal frame $(\qo_{ab}^\prime, [\ello^{\prime\,a}]) = \alpha^2 \qo_{ab},\, [\alpha^{-1}\ello^{\prime\,a}])$, the vector space generated by linear combinations of dilations and supertranslations remains the same. 
\footnote{If we change the fiducial pair $(\ello^a, \vo) \to (c^{-1} \ello^a,\, c\vo + f(x^A))$, the sub-algebra $\mathfrak{v}$ of vertical vector fields remains invariant, and the dilation part of $\xi^a$ also remains invariant but the supertranslation part changes via $s(x^A) \to c\, s(x^A) + f(x^A)$.}
However, `pure' Lorentz transformations with respect to one conformal frame are no longer so with respect to another. Again this situation is completely analogous to that at $\scrip$.
 
We will use the expressions (\ref{QN3}) and (\ref{Q}) of charges and the decompositions (\ref{xi2}) and (\ref{xi4}) of symmetry vector fields to define dilation energy, supermomenta, and the Lorentz angular momentum. Note that in the derivation of (\ref{QN3}) and (\ref{Q}), we only assumed that $g\in \ps$; thus $\N$ is \emph{not} required to be an NEH. We will use these expressions for NEHs in this section, and for perturbed NEHs in Sec.~\ref{s4}.\\

\emph{Remark:} While the NEH symmetry group $\G$ arose \cite{akkl1} as the group preserving the universal structure provided by the pairs $(\qo_{ab}, [\ello^a])$ on $\H$, as we remarked in Sec.~\ref{s1}, on any given `concrete' NEH $\H$, the symmetry vector fields $\xi^a$ can be also be identified by their action on pairs $(q_{ab}, \, [\ellb^a])$, directly induced on $\H$ by the space-time metric $g_{ab}$, via:
\be \label{xi3}
\Lie_\xi q_{ab} = 2\phi\, q_{ab} \qquad {\rm and} \qquad \Lie_\xi \ellb^a = -(\phi + k)\, \ellb^a \ee
This characterization of symmetries provides an equivalent but alternate decomposition of $\xi^a$. Given a concrete NEH, then, let us choose an $\ellb^a \in [\ellb^a]$, fix an affine parameter $\b{v}$ of $\ellb^a$, and consider $\b{v}={\rm const}$ foliation. Then, Eq.~(\ref{xi3}) implies that $\xi^a$ can be uniquely decomposed as
\be \label{xi4}
\xi^a = d^a + b^a = d^a + (S^a+L^a) = k \b{v} \ellb^a + \big(\b{s} \ellb^a + (\phi \ellb^a + \b{H}^a)\big)\, , \ee
where $b^a$ is a BMS vector field that is a sum of a supertranslation $S^a$ and a Lorentz transformation  $L^a = \phi \ellb^a + \b{H}^a$. The `horizontal part' $\b{H}^a$ is tangential to the $\b{v}={\rm const}$ cross-sections; is Lie-dragged by $\ellb^a$,\, $\Lie_{\ellb} \b{H}^a =0$;\, and is a conformal Killing field of $q_{ab}$ satisfying $\Lie_{\b{H}} q_{ab} = 2\phi q_{ab}$. Thus, $2\phi = D_a \b{H}^a$ where $D_a$ is the derivative operator on the $\b{v}={\rm const}$ cross-sections.\\

\subsection{Simplifications on $\psneh$}
\label{s3.2}
 
For notational convenience, let us decompose a general symmetry vector field $\xi^a$ of (\ref{xi2}) in to \emph{its} vertical and horizontal parts:
\be \xi^a = V^a_{(\xi)} + H^a_{(\xi)}, \quad {\rm with}\quad  V^a_{(\xi)} = (\vo\,(k+\phio)\, +\, \so) \ello^a \quad {\rm and}\quad  H^a_{(\xi)} = \epsilono^{ab} \Do_b \chio\ + \qo^{ab} \Do_b \phio\, . \ee 
Thus, in terms of the extension (\ref{X1}) $X^a$ of $\xi^a$, we have: $f_1 = k +\phio,\, f_2 = \so$ and $H^A = H_{(\xi)}^a\, \nabla_a X^A$ (in the chart $(v,r,x^A)$ determined by $\ello^a$ and its affine parameter $\vo$ on $\H$.)\smallskip

Let us now restrict ourselves to metrics $g\in \psneh$ so that $\N$ is an NEH; thus  
$\theta_{(\ell)}\, \=\,0$ for all null normals. Hence, the second term in the expression (\ref{Q}) of the charge $Q_{\xi}[C]$ vanishes and we are left just with the Noether Charge. For dilations and supertranslations, $\phi =0$ and the horizontal part $H^a_{(\xi)}$ also vanishes. Therefore, the corresponding charges are given by
\be \label{charge1} Q^{(0)}_d\,[C] = \f{1}{8\pi G}\, k\,  A [C]  \qquad {\rm and} \qquad Q^{(0)}_S\, [C] =0\, , \ee
where $A\,[C]$ is the area of the 2-sphere $C$, and where the superscript $(0)$ in $Q^{(0)}$ is a reminder that this charge refers to an \emph{unperturbed} NEH. For rotations and boosts we have:
\be \label{charge2} Q^{(0)}_R\, [C]\,=\, -\f{1}{16\pi G}\, \oint_C R^c \beta_c\, \epsilon_{ab} \qquad {\rm and} \qquad
 Q^{(0)}_B\, [C]\,=\, -\f{1}{16\pi G}\, \oint_C \big(2\phio\, -\, {B}^c \beta_c \big)\,\epsilon_{ab}\, . \ee 

Finally, it is instructive to use the second decomposition (\ref{xi4}) of $\xi^a$ and obtain the corresponding charges without any reference to pairs $(\qo_{ab}, [\ello^a])$. Let us use for the null normal in (\ref{QN2}) an $\ellb^a$ from the canonical equivalence class $[\ellb^a]$ on $\H$ and simplify: 
\ba \label{charge3}
Q_\xi\, [C] &=& - \f{1}{8\pi G}\,\oint_C \epsilon_{ab}\,\, \ellb^c \b{n}^d\, \nabla_{c} X_{d}
= - \f{1}{8\pi G}\,\oint_C \epsilon_{ab}\,\, \big( - (k +\phi) + \xi^c \bar{n}_d \nabla_c \ellb^d \big)\nonumber\\
 &=& \f{1}{8\pi G}\,\oint_C \epsilon_{ab}\,\, \big((k + \phi) + \b{H}^c \omega_c\big)\, =\, \f{1}{8\pi G}\,\oint_C \epsilon_{ab}\,\, \big((k +\textstyle\f{1}{2} D_c \b{H}^c) + \b{H}^c \omega_c\big). \,\,\,\,
\ea
Here: (i) in the first step we have removed anti-symmetrization over $c$ and $d$ using the fact that the vector field $X^a$ satisfies $\ell^b \Lie_X g_{ab}\, \=\,0$ for all null normals (since for any given $\ell^b$, the 1-form $\ellb_a \,:\= \, g_{ab}\ell^a $ is the same for all $g \in\ps$); (ii) in the second step, we used the fact that $X^a$ is tangential to $\H$ and satisfies $\Lie_{X} \ellb^a \,\=\, -(\phi + k) \ellb^a$; and, (iii)  in the third step the fact that $\pb{\nabla_{a}} \ellb^{\,b} = \omega_a \ellb^{\,b}$, where $\omega_a$ is the (divergence-free) rotational 1-form; and, (iv) in the fourth step, the fact that $\b{H}^a$ is a conformal Killing field of the metric $q_{ab}$ with $\Lie_{\b{H}} q_{ab} = 2\phi\, q_{ab}$. In this derivation, we used the null normal $\ellb^a$ but we did not have to introduce coordinates. On the other hand, in expressions (\ref{charge2}) of rotation and boost charges we introduced the chart $(v,r,x^A)$ adapted to the null normal $\ello^a$. 
If $\qo_{ab}  = \psi^2 q_{ab}$, it is easy to verify that $\phio = \phi + \Lie_{\b{H}}\ln \psi$ and $\beta_a = -2 (\omega_a + D_a \ln \psi)$. Therefore the two sets of charges agree, as they must. Thus (\ref{charge3}) encodes the dilation, supermomentum, rotation and the boost charges we found above using a conformal frame $(\qo_{ab}, [\ello^a])$. The expression (\ref{charge3}) has the advantage that it makes it manifest that the final values of charges depend only on $g\in \psneh$ and the symmetry vector field $\xi^a$ and are independent of the auxiliary structures such as the choice of a pair $(\qo_{ab}, [\ello^a])$.\\

\emph{Remarks}:

(1) In the quasi-local horizon framework, angular momentum is unambiguously defined for axisymmetric NEH geometries and is then given by $J_{\H} = - \f{1}{8\pi G} \oint_C \omega_a R^a$ where $R^a$ is the generator of rotational symmetry  \cite{abl2,akrev}. It follows from Eqs.~(\ref{charge2}) and/or (\ref{charge3}) that $J_{\H} = - Q_{R} [C]$. Consequently, on general NEHs where the geometry is not necessarily axisymmetric, we are led to interpret $J^{(0)}_{R}[C] := -Q^{(0)}_{R} [C]$  as the component of the NEH angular momentum corresponding to the NEH symmetry $R^a$. 

(2) As we noted in footnote \ref{fn5}, each concrete NEH is equipped with a canonical unit round metric $\qo_{ab}$ for which the `area dipole' vanishes \cite{akkl1}. If one uses this $\qo_{ab}$ to carry out the decomposition of $\xi^a$ into various parts, the first term involving $\phi$ in the boost angular momentum in (\ref{charge2}) vanishes because $\phi$ is a linear combination of the first three spherical harmonics. However, this specific canonical round metric $\qo_{ab}$ is tied to physical metric $q_{ab}$; it is not universal. \smallskip

Are the specific values of charges we found compatible with our physical intuition? Recall that on the full phase space $\ps$ where $\N$ is only required to be a null boundary, this is not the case because, for example, the charge and flux associated with the dilation symmetry on (possibly, a portion of) the null cone of \emph{Minkowski space} is non-zero and time-dependent because $\oint_C \theta_{(\ell)}\, d^a\, \epsilon_{abc}$ fails to vanish and varies from one cross-section to another. Do such physically spurious features persist even though we have restricted ourselves to 2-sphere cross-sections $C$ of an NEH? 

Let us first consider fluxes. Recall first that NEHs $\H$ are more general than the Killing horizons. For example, in the asymptotically de Sitter context the Kastor-Traschen solutions represent dynamical, multi-black hole space-times that admit NEHs which are not Killing horizons \cite{kt,kt1, kt2}. In the zero cosmological context, the Robinson-Trautman solution admits an NEH which is not a Killing horizon; in fact
every neighborhood of the NEH contains radiation \cite{pc}. Nonetheless, the area of cross-sections of NEHs does not change. The mathematical reason is that the expansion $\theta_{(\ell)}$ vanishes and the physical reason is that the definition of NEHs ensures that even though there may be radiation arbitrarily close to $\H$, none falls into $\H$. Therefore,  one expects that fluxes of physical quantities across $\H$ should vanish. Is this expectation borne out? Since $\epsilon_{ab}$ and $\omega_a$ are Lie-dragged by every null normal $\ell^a$ and also transverse to it, it follows that the values of these charges are independent of the choice of the cross-section $C$. Therefore, all fluxes vanish identically, just as one would expect. One can also verify this directly using the expression (\ref{flux}) of fluxes.  

What about the charges? Let us begin with angular momentum. Consider axisymmetric solutions $g_{ab}$ in $\psneh$ that are asymptotically flat at spatial infinity, in which the axial Killing field $\bar{R}^a$ is tangential to $\H$. As we noted above, on an NEH, the charge $Q^{(0)}_\xi [C]$ associated with any horizon symmetry $\xi^a$ is given by the Noether charge of \emph{the extension} $X^a$ of this horizon symmetry. Now, a key feature of our extensions $X^a$ is that if $\xi^a$ is the restriction to $\H$ of any space-time Killing vector, then $X^a$ agrees with that Killing field (to the order that enters the expression of the Noether charge; see Appendix \ref{a2}). Therefore, in the axisymmetric space-times now under consideration, the charge $Q^{(0)}_R[C]$ associated with the NEH symmetry $R^a$ induced by the space-time Killing field $\bar{R}^a$ satisfies 
\be Q^{(0)}_R\,[C]\, = \,Q_{\bar{R}}^{\rm N} \,[C]\, \equiv \, - \f{1}{16\pi G}\, \oint_C \epsilon_{abcd} \nabla^c \bar{R}^d \, . \ee
In vacuum space-times, $Q_{\bar{R}}^{\rm N} \,[C]$ is precisely negative of the Komar integral for angular momentum associated with the Killing field $\bar{R}^a$. The value of the Komar integral remains remains unchanged if we continuously deform the 2-sphere $C$ to another 2-sphere $C^\prime$. Furthermore, in the limit that $C$ approaches spatial infinity, it yields $J^{\rm ADM}_{\bar{R}}$. Thus, our interpretation that $J^{(0)}_R\,[C] = - Q^{(0)}_R\,[C]$ is the horizon angular momentum (associated with the rotation symmetry $R^a$) implies $J^{(0)}_R\,[C] = J^{\rm ADM}_{\bar{R}}$ for space-times that admit a global axial Killing field $\bar{R}^a$, just as one would hope. 

What about translations? Initially it is rather surprising that the 4-dimensional translation subgroup of the BMS group is now extended to a 5 dimensional group due to the presence of the dilation in $\G$. Although, as explained in detail in \cite{akkl1}, this extension is inevitable, does it lead to any physically uncomfortable results with respect to charges? For example, we just found that all supermomentum charges vanishes identically if $\N$ is an NEH. Does this result not lead to a tension in the case when $\H$ is a Killing horizon, given that the ADM charge associated with the time translation Killing field is non-zero? To shed light on such issues we will now evaluate the charges in a few simple yet physically interesting examples. We will find that there is no tension at all.

\subsection{Examples}
\label{s3.3}

In the interplay between time-translation Killing fields in space-time and the horizon charges one encounters two subtleties. To illustrate the first, let us begin with the Schwarzschild metric. Then, our internal boundary $\N$ is a Killing horizon, with surface gravity $\kappa = 1/4GM$ associated with the time translation Killing field $\b{t}^a$ that is normalized to be unit at infinity. For simplicity, let us use the Eddington-Finkelstein coordinates that are adapted to the Killing symmetry, so that:
\be \label{sch} g_{ab} \rmd x^a \rmd x^b = -\left( 1 - \frac{2GM}{\b{r}}\right) \rmd \b{v}^2 + 2\rmd \b{v} \rmd \b{r} + \b{r}^2 \rmd \theta^2 + \b{r}^2 \sin^2 \theta \rmd \varphi^2 \, .\ee
Since the horizon is located at $\b{r}=2GM$, the coordinate $r$ in Eq. (\ref{metric}) is given by $r = \b{r} -2GM$. More importantly, in a neighborhood of the horizon, the metric (\ref{sch}) is not in the form given in Eq.(\ref{metric}) because in the Eddington Finkelstein coordinates $\partial_{\bar{v}} = \b{t}^a \partial_a$, whence its restriction to $\H$ does not coincide with the affinely parametrized null geodesic vector field $\partial_v$. On the NEH, this Killing field can be expressed in the $v,r, \theta,\varphi$ chart as $\partial_{\bar{v}} = \f{1}{4GM}\, v\, \partial_v$. \emph{Thus, it is a dilation symmetry $d^a = k\, v\, \partial_v$ with $k = \f{1}{4G M}$.} Therefore, the associated charge (\ref{cfpcharge}) is
\be Q^{(0)}_{d} [C] = \f{1}{32\pi\,GM}\,\, A [C] \,= \,\f{M}{2}\, . \ee
That the result is $M/2$ rather than $M$ may seem puzzling at first. However
further thought shows that this is precisely what one should expect because: (i) As we noted above, $Q^{(0)}_{d} [C]$ equals the Noether charge associated with the extension $X^a$ of $d^a$, and $X^a$ agrees with the Killing field $\b{t}^a$ to the desired order;\, and,\, (ii) It is well-known that the Noether charge associated with the time translation Killing field in any stationary space-time is half the ADM mass. Thus we are led to interpret $2 Q^{(0)}_{d} [C] =: E^{(0)}_{\b{t}} [C] $ as the NEH energy associated with the Killing symmetry $\b{t}^a$. Note that because restriction of the Killing field $\b{t}^a$ to $\H$ is a dilation, there is no tension with the fact that all supertranslation charges vanish. Indeed, there is no Killing field in the Schwarzschild space-time whose restriction of the horizon yields a supertranslation. 

To bring out the second subtlety associated with time-translation symmetries, let us  
turn to the Kerr solution. Denote by $\b{t}^a$ the stationary Killing field that has unit norm at infinity and by $\b{R}^a$ the rotational Killing field whose affine parameter takes values in $(0, 2\pi)$. Mass $M$ is associated with $\b{t}^a$ and angular momentum $J$ associated with $\bar{R}^a$. Does our interpretation of the horizon charges $Q^{(0)}_{d} [C]$ and $Q^{0}_R [C]$ reproduce $M$ and $J$ correctly? NEH supertranslations played no role in the Schwarzschild space-time. Do they perhaps play a role now? If so, is there not a tension with the fact that all charges $Q^{(0)}_{S} [C]$ vanish? Since these issues can cause unease, we will carry out explicit calculations to make the results fully transparent.

Let us begin with the interplay between the dilation/supertranslation symmetries of $\H$ and the space-time Killing fields of Kerr space-time. The problem neatly divides into two parts because: (i) in the non-extremal case there is \emph{no} Killing field  --i.e., no constant linear combination of $\b{t}^a$ and $R^a$-- that is a supertranslation on $\H$, while there is a unique linear combination (modulo a constant rescaling) that it a dilation; and, (ii) in the extremal case there is \emph{no} Killing field that is a dilation on $\H$, while there is a unique linear combination (modulo a constant rescaling) that it a supertranslation.

Let us begin with the non-extremal case $J < GM^2$. Then, our null boundary $\H$ is a \emph{Killing horizon} for the following linear combination of the two Killing fields
\be \b{K}^a = \b{t}^a + \Omega_\H\, \b{R}^a, \quad {\rm with} \quad \Omega_\H  = \frac{J/G}{2MG \big(M^2 + \sqrt{M^4 - J^2/G^2}\big)}\, . \ee
As is well-known, $\Omega_H$ has the interpretation of the angular velocity of the horizon which vanishes in the Schwarzschild limit and equals $1/2GM$ in the extremal case $J = GM^2$. The Killing field $\b{K}^a$ becomes a dilation symmetry on $\H$, given by
\be \b{K}^a|_\H = d^a = k\, v \ell^a \quad {\rm where} \quad k = \frac{\sqrt{M^4 - J^2/G^2}}{2GM(M^2 + \sqrt{M^4 - J^2/G^2})}\, . \ee
$k$ is the surface gravity that the Killing field $\b{K}^a$  endows on the horizon, and $v$ is the affine parameter of the geodesic vector field $\ell^a \in [\ell^a]$. Now, from Eq.~(\ref{charge1}) we have $Q^{(0)}_d [C] = (k\,A[C])/(8\pi G)$. While in the Schwarzschild case this expression just gave us $M/2$, now is is a rather complicated function of $M$ and $J$:
\be A_\H =  8\pi G^2 \big( M^2 + \sqrt{M^4 - J^2/G^2}\big) \ee
However, using the expressions of $\Omega_\H$ and $k$, we can rewrite $Q^{(0)}_d [C]$  as the expected linear combination of the NEH energy and angular momentum, associated with the NEH symmetries $\b{t}^a$ and $\b{R}^a$ respectively
\be \label{noether_dilation_kerr} 
Q^{(0)}_d [C] \, := \, \frac{k A_\H}{8\pi G}\, =\, \frac{M}{2}\, - \, \Omega_\H\, J \, =\, Q^{(0)}_{\b{t}} + \Omega_\H Q^{(0)}_{\b{R}}  . \ee
Thus, although the initial expression (\ref{charge1}) of the dilation charge\, $Q^{(0)}_d [C]$\, seems rather opaque, it has precisely the value one would expect from the fact that $d^a \,\=\, (\b{t}^a + \Omega_H \b{R}^a)$.

Finally, let us consider the extremal Kerr solution. In this case $\H$ is a Killing horizon with zero surface gravity, whence the restriction of the Killing field $\b{K}^a$ to $\H$ is a \emph{supertranslation} $S^a$ \emph{rather than a dilation}. But we saw in Eq. (\ref{charge1}) that charges $Q^{(0)}_S [C]$ associated with all supertranslations vanish identically on the NEH! Is there then a tension with the fact that the mass and the angular momentum of this solution are non-zero? Again, the answer is in the negative. Using the fact that $ J = GM^2$ and $\Omega_\H = 1/2GM$ in the extremal case, $S^a \, \=\, \b{t}^a + \Omega_H\, \b{R}^a$ 
yields:
 \be Q^{(0)}_{S}\, [C]\, = \, Q^{(0)}_{(\b{t})}\, [C]\, + Q^{(0)}_{(\b{R})}\, [C]\, 
  = \f{M}{2} - \f{M}{2} =0\, . \ee
Thus, even though neither $M$ nor $J$ vanish, the horizon charge associated with the supertranslation $S^a$ induced by $\b{K}^a$ on $\H$ does vanish simply because the linear combination of the two Killing fields for which $\H$ is a Killing horizon is very specific. There is no tension between the fact that the mass and angular momentum of the extremal Kerr are non-zero and vanishing of all supertranslation charges. \\

We will conclude with a few remarks.\smallskip

(i) Since one arrives at the expressions of charges using phase space methods, one has direct control only on variations of charges. Thus, to begin with one only knows
\be \delta Q_\xi [C]\, =\,  \delta Q^{\rm N}_\xi\, [C] - \f{1}{8\pi G}\,\, \delta\,\oint_C \theta_{(\ell)}\, \xi^a \, \epsilon_{abc} \, \ee
for all permissible tangent vectors $\delta g$ at the solution $g_{ab}$ under consideration. To arrive at the expression of the charges themselves, one needs to fix the integration constant on the phase space. Normally, one eliminates this freedom by choosing a preferred background, typically  Minkowski space, and specifying the values of charges at that solution. Minkowski space is not in our $\psneh$. Instead, we can consider the 1-parameter family of Schwarzschild solutions and demand that all charges should vanish in the limit $M \to 0$. This requirement eliminates the freedom to add (the $\xi$-dependent) constants to the charges, and leads us to the definition (\ref{cfpcharge}).

(ii) Since we have used some results from \cite{cfp}, to avoid potential confusion, let us note points of difference between that framework and ours. First, as we have already emphasized, we are primarily interested in the sub-manifold $\psneh$ of the covariant phase space $\ps$ considered in \cite{cfp}; thus our charges and fluxes refer only to non-expanding horizons and perturbations thereof. The second point is a bit more subtle. We have restricted our null normals $\lcirc^a$ and $\ello^a$ to $\H$ to be affinely parametrized geodesic vector fields. Therefore, in the notation of \cite{cfp}, their $\kappa$ takes the value zero in our analysis. Nonetheless, when we analyzed the symmetry vector fields that preserve our universal structure, we were naturally led --in fact, forced-- to admit dilations $d^a = k\, \vo \ello^a$ in our symmetry Lie algebra $\g$. The acceleration of these vector fields on any given `concrete horizon' is non-zero and given by $k$ (since $\ello^a$ is a geodesic vector field on all `concrete' horizons in our phase space $\ps$). Thus, in our framework, non-zero acceleration is associated \emph{only with symmetries; not with the null normals that horizons come equipped with}. (We labelled the the dilations with the letter $k$ rather than $\kappa$ to avoid potential confusion with formulas between \cite{cfp} and our work.) 

(iii) As in \cite{cfp}, our symmetry vector fields $\xi^a$ are `kinematical', i.e, the same for all phase space points $g \in \psneh$. In particular, then, any given dilation $d^a = k\, \vo\ello^a$ comes with a fixed positive number $k$, which represents the acceleration of the null vector field $d^a$ in any concrete space-time in $\psneh$. Now, the phase space $\ps$ admits, e.g., Schwarzschild solutions with any positive mass $M$, each with a static Killing field $\b{t}^a$ normalized to be a unit time translation at infinity. While this $\b{t}^a$ is a dilation on $\H$ for all values of $M$, its acceleration on $\H$ --or  surface gravity-- is $1/4M$ and thus varies from one solution to another. Therefore, \emph{there is no NEH symmetry $\xi^a$ to which static Killing fields of \emph{all} Schwarzschild solutions can tend to on $\H$}. As a result, as in \cite{cfp}, in our framework there is no NEH symmetry $\xi^a$ such that $Q_{\xi} = M/2$ for \emph{all} Schwarzschild metrics. On the other hand, this was possible in the covariant phase framework of \cite{afk,abl2} because that framework allowed `live' vector fields $\xi^a$ as symmetries, i.e. $\xi^a$ themselves could vary in a streamlined manner from one phase space point to another. (Note that this is the analog of `live' lapse and shift fields used routinely on the canonical phase space.) A generalization of the phase space framework used in this paper that allows the symmetry vector fields to be `live' may lead to a new definition of charges such that $Q^{\rm new}_\xi$ is the ADM mass for the entire Kerr family, for suitably chosen live symmetry vector field $\xi^a$. It is worth exploring this issue further.%
\footnote{Interestingly, this subtlety  does not arise for rotations because there is a universal normalization for the rotational symmetry vectors $R^a$: They have closed orbits and normalized so that their affine parameter runs from $0$ to $2\pi$, irrespective of the angular momentum content of space-time. Consequently, our charge $Q_{R}$ yields the correct angular momentum $J$ in any axisymmetric, asymptotically flat space-time.}

\section{Charges and fluxes on perturbed NEHs}
\label{s4}

This section is divided into two parts. In the first, we collect results on first and second order linearization and in the second part we discuss charges and fluxes.

\subsection{First and Second Order Perturbations}
\label{s4.1}

Let us now consider a 1-parameter family of solutions $g(\lambda) \in \ps$ (that depends smoothly on $\lambda$) such that $\gcirc_{ab} := g_{ab}(\lambda)|_{\lambda=0}$\, is in $\psneh$\, (so that $\N$ is an NEH $\H$ for $\gcirc_{ab}$). We can carry out a Taylor expansion in $\lambda$:
\ba \label{taylor} g_{ab} (\lambda) &=& \gcirc_{ab} \,+\, \lambda\, \f{\rmd g_{ab}(\lambda)}{\rmd \lambda}|_{\lambda =0}\, +\, \f{\lambda^2}{2} \,\f{\rmd^2\, g_{ab}(\lambda)}{\rmd \lambda^2}|_{\lambda =0} + \ldots\nonumber\\
&=:& \gcirc_{ab} + \lambda\, \hone_{ab}\, +\,  \f{\lambda^2}{2}\, \htwo_{ab} + \ldots \ea
so that $\hone_{ab}$ is a first order perturbation, $\htwo_{ab}$, a second order perturbation, etc, on the background solution $\gcirc_{ab}$. As usual, indices will be raised and lowered using the background metric $\gcirc_{ab}$. We will use the structure on the boundary $\N$ made available by the fact that it is an NEH of the background metric $\gcirc_{ab}$, and express the symmetry vector fields $\xi^a$ using that structure as in Sec.~\ref{s3}. We will also use the fact that, while
$\N$ is perturbed and no longer an NEH due to $\hone_{ab}$ and $\htwo_{ab}$, the perturbations are controlled by the fact that $\hone_{ab}$ satisfies the linearized Einstein's equations on the $\gcirc_{ab}$ background as well as the asymptotic conditions (in the distant future) introduced in Sec.~\ref{s2}, and $\htwo_{ab}$ satisfies the second order linearized equation. We will primarily use the implications of these equations on the Raychaudhuri equation that governs properties of null geodesics on $\N$. 
\smallskip

On the NEH of the background metric $\gcirc_{ab}$, the expansion and the shear of these geodesics vanish: 
\be \thetacirc\, =0 \qquad {\rm and} \qquad \sigmacirc_{ab}\, =0\, . \ee
Furthermore, as we noted in Sec.~\ref{s2}, because of our choice of the phase space $\ps$ the expansion continues to vanish at the first order but the shear does not.  Let us use the notation $F^\prime(g) = \f{\rmd}{\rmd \lambda}\, F(g(\lambda))|_{\lambda=0}$, and similarly for double prime. Then to first order we have:
\be \label{firstorder1}\epsilonone_{ab} = {\textstyle{\f{1}{2}}}\,(\qcirc^{cd}\,\,\hone_{cd})\,\epsiloncirc_{ab} \,=:\,{\textstyle{\f{1}{2}}}\,\, (\ubhone)\, \, \epsiloncirc_{ab} \qquad{\rm and} \qquad 
\thetaone\, =\, {\textstyle{\f{1}{2}}}\,\, \dot{\ubhone} \,=\, 0\ee
where the `dot' denotes derivative w.r.t. $v$,\, and,
\be \label{firstorder2}\sigmaone_{ab}\, =\, {\textstyle{\f{1}{2}}}\,(\qcirc_a{}^c)\,\, (\qcirc_b{}^d) \,\,\Lie_{\ell}\, (\hone_{bd}) \qquad{\rm or,}\qquad 
\sigmaone_{AB} = {\textstyle{}\f{1}{2}}\, (\honedot_{AB})\, .\ee 
To second order, we will only need $\epsilontwo_{ab}$ and $\thetatwo$, which are given by 
\ba \label{secondorder}\epsilontwo_{ab}\, &=&\, {\textstyle{\f{1}{2}}}\,  \big(\ubhtwo\, \ -\, \hone_{ab}\, \hone_{cd}\, \qcirc^{ac}\, \qcirc^{bd}\, +\, {\textstyle{\f{1}{2}}}\, (\ubhone)^2 \big)\,\,\epsiloncirc_{ab} 
\nonumber\\
\thetatwo\, &=& \, {\textstyle{\f{1}{2}}}\,\dot\ubhtwo  \,- \,  ({\honedot}_{ab})\,( \hone_{cd})\,\, \qcirc^{ac}\, \qcirc^{bd}\, .\ea
Because of the asymptotic conditions in the distant future we imposed on solutions $g\in \ps$\, (and the Raychaudhuri and Einstein equations), the fields $\thetatwo$ and $\sigmaone_{AB}$ fall-off at least as fast as $1/|v|^{1+\epsilon}$ as $|v| \to \infty$.
\smallskip

In the next subsection we will evaluate the first and second order corrections to the charges $Q^{(0)}_\xi$. We will restrict ourselves to space-time metrics $g_{ab}(\lambda) \in \ps$ of the form (\ref{taylor}). Since $\N$ is an unperturbed NEH for the background solutions $g_{ab}(0)$, we can use the symmetry group $\G$ of the unperturbed NEH to compute fluxes and charges associated with not only $g_{ab}(0)$, but also those associated with the first and second order perturbations thereon. This is analogous to the fact that given a stationary space-time, we can use the stationary Killing field to speak of energy associated with that solution as well as that carried by perturbations about that space-time.

Let us fix a symmetry vector field $\xi^a$ and its extension $X^a$. Neither has $\lambda$-dependence as we move along our 1-parameter family of metrics.  However\, $\nabla^a,\, \epsilon_{abcd}, \epsilon_{abc}$ and $\theta_{(\ell)}$ in the integrand of charge integrals  acquire a $\lambda$-dependence. Therefore, using (\ref{Q}) (or (\ref{QN2}) or (\ref{QN3})) that hold for any $g\in \ps$, we obtain a 1-parameter family $ \big(Q_\xi\, [C]\big) (\lambda)$ of charges. By evaluating the charge at $\lambda=0$, and taking its first two derivatives with respect to $\lambda$ at $\lambda=0$, we obtain: \\ 
\indent (i) Charges $Q_\xi^{(0)}\, [C]$, evaluated in the background solution $\gcirc_{ab}$. We already discussed these charges in Sec.~\ref{s3}. However, in the notation we just introduced to handle perturbations, the fields\, $\omega_c,\, \epsilon_{ab}$\, and the area\, $A\,[C]$\, that appear in the expressions of the NEH charges should be replaced by \,${}^\circ\!\omega_c,\, \epsiloncirc_{ab}$ and ${}^\circ\!{A}\,[C]$,\, as they refer to the background metric $\gcirc_{ab}$.\\
\indent (ii) The first order perturbation $Q_\xi^{(1)}\, [C]$ of this charge; and,  \\
\indent (iii) The second order perturbation $Q_\xi^{(2)}\, [C]$ of this charge. \\
  
\subsection{First and Second order Corrections to Charges and Fluxes}
\label{s4.2}
 
As in Sec.~\ref{s3}, let us discuss divide the discussion of the perturbed charges and fluxes using the decomposition of $\xi^a$ into various parts. Then the expression (\ref{Q}) of charges on the full phase space $\ps$ implies that the first order corrections to dilation and supertranslation charges are given by:
\be \label{nehcharge3} Q_{d}^{(1)}\, [C] = \f{k}{8\pi G}\, A^\prime\, [C] = \f{k}{16\pi G}\, \oint_C \ubhone\, \epsiloncirc_{ab} \qquad {\rm and} \qquad  Q_{S}^{(1)} = 0\, .\ee
where we have used the fact that $\epsilon^\prime_{ab} = \f{1}{2}\,(\qcirc^{cd}\,\,\hone_{cd})\,\epsiloncirc_{ab} \,=:\,\f{1}{2}\,\, (\ubhone)\, \, \epsiloncirc_{ab}$
and the fact that the expansion vanishes to first order, i.e., $\theta_{(\ell)}^\prime =0$. Next, let us consider rotations and boosts. From (\ref{Q}) we have: 
\ba \label{nehcharge4}Q_{R}^{(1)}\, [C] &=& -\f{1}{16\pi G}\oint_C R^c (\beta^\prime_c + {\textstyle{\f{1}{2}}}\,\, \ubhone\, \beta_c)\,\, \epsiloncirc_{ab} \nonumber\\
Q_{B}^{(1)}\, [C] &=& \f{1}{16\pi G}\oint_C \big(2\phio\,\, \ubhone -\tilde{B}^c (\beta^\prime_c   +{\textstyle{\f{1}{2}}}\,\,\ubhone\, \beta_c)\big)\, \epsiloncirc_{ab}\, , 
\ea
where $\t{B}^a$ is the horizontal part of the boost vector field $B^a$. Thus, while for the dilation and supertranslations the perturbed charges are expressed in terms only of perturbations of the 2-metric $q_{ab}$, for rotations and boosts, they also involve perturbations of $\beta_a$ --the radial derivative of perturbations of the `$g_{vA}$ part' of the metric-- just as one would expect from the expression (\ref{charge2}) of these charges on the NEH. Finally, note that $\Lie_{\ell}\, \ubhone\, \=\,0$ on $\N$ because $\thetaone \,\=\,0$ there and $\Lie_{\ell} \beta_c$ also vanishes by Eq.~(\ref{background}). Under a generic perturbation, $\Lie_{\ell} \beta^\prime_C$ and $\beta^\prime_c$ do not vanish on $\N$. However, the Linearized Einstein's equations $R^\prime_{ab}\, \ell^a\, \qcirc^{bc}\, \=\,0$ and the fact that $R^a$ and $\t{B}^a$ are conformal Killing fields of the physical metric imply that the integrals of $R^a (\Lie_{\ell}\,\beta^\prime_a)$ and $\t{B}^a (\Lie_{\ell}\,\beta^\prime_a)$ over the cross-section $C$ vanish. Hence to first order all fluxes vanish, as as they do for perturbations of stationary space-times at $\scrip$. Towards the end of this subsection we will provide an alternate proof of vanishing of all fluxes to first order that does not refer to coordinates or the metric component $\beta_A$.
\smallskip

Next, let us consider second order corrections to charges and fluxes. Taking the second $\lambda$-derivative of (\ref{Q}) and evaluating at $\lambda=0$ we have,
for the dilation and supertranslations, 
\be \label{nehcharge5} Q_{d}^{(2)}\, [C] = \f{1}{8\pi G}\,\oint_C   k\,\, \epsilontwo_{ab} - \theta_{(\ell)}^{\prime\prime}\,d^c\, \epsiloncirc_{abc}\qquad {\rm and} \qquad
Q_{S}^{(2)}\, [C] = -\f{1}{8\pi G}\,\oint_C  \thetatwo\,S^c\,\,\epsiloncirc_{cab}
\ee
where expressions of $\thetatwo$ and $\epsilontwo$ in terms of the metric perturbations are given in (\ref{secondorder}). For rotations and boosts, 
because $H^c \epsilon_{c\,\pb{ab}} \, \=\, 0$ when the indices $ab$ are pulled back to the cross-section $C$ to which the vector field $H^c$ is tangential, the second term in (\ref{Q}) does not contribute and we obtain
\ba \label{nehcharge6} Q_{R}^{(2)}\, [C] &=&  -\f{1}{16\pi G}\,\oint_C R^c\,\big(\beta_c\, \epsilon_{ab})^{\prime\prime} \nonumber\\
   Q_{B}^{(2)}\, [C] &=& \f{1}{16\pi G}\,\oint_C  {2\phio\, \epsilon^{\prime\prime}_{ab}}\,-\, \tilde{B}^c\, \big(\beta_c\, \epsilon_{ab})^{\prime\prime}\,   \ea
where $\t{B}^a$ is again the horizontal part of the boost $B^a$ vector field.

As one would expect from the behavior of perturbations at $\scrip$, the second order fluxes associated with these charges are non-zero. They provide the balance laws for the perturbed charges. For the dilation and supertranslations, fluxes can be readily obtained using the Raychaudhuri equation:
\ba \label{flux1}\F^{(2)}_{d} [\N_{1,2}]\, &=&\, -\f{1}{4\pi G}\, \int_{\N_{1,2}}|\sigmaone_{mn}|^2\,\, (d^d n_{d})\, \epsiloncirc_{abc}\,\, =\,\, -\f{1}{16\pi G}\, \int_{\N_{1,2}} |\dot\hone_{mn}|^2\, (d^d n_d)\, \epsiloncirc_{abc} \nonumber\\
\F^{(2)}_{S} [\N_{1,2}]\, &=&\, -\f{1}{4\pi G}\, \int_{\N_{1,2}} |\sigmaone_{mn}|^2\, (S^d n_{d})\,\epsiloncirc_{abc}\,\, =\,\, -\f{1}{16\pi G}\, \int_{\N_{1,2}} |\dot\hone_{mn}|^2\, (S^d n_{d})\, \epsiloncirc_{abc}\,.\nonumber\\  \ea
Here, the dot denotes Lie derivative w.r.t. $\ell^a$,\, we have set $|\sigmaone_{mn}|^2 = \sigmaone_{mn}\, \sigmaone_{st}\,\,  \qcirc^{ms}\,\, \qcirc^{nt}$ and similarly for $|\dot\hone_{ab}|^2$, and $n^a$ is any vector field such that $g_{ab} \ell^a n^b\, \=\, -1$. The expression $\F^{(2)}_{d} [\N_{1,2}]\,$ generalizes the formula for energy flux carried by gravitational waves across a de Sitter horizon given in \cite{kl2020}.

In the calculation of $\F^{(2)}_{d}$, there \emph{are} terms with indeterminate sign in the intermediate steps, but they all cancel out and the flux is non-negative (since $d^a n_{a}$ is negative). It vanishes if and only if the perturbation $\hone_{AB}$ of the 2-metric is time-independent. An example of such a perturbation  is provided by a $\hone_{ab}$ that changes just the mass and/or angular momentum along the Kerr family. If $\hone_{ab}$ is time dependent, the charge $Q_d^{(2)} [C]$ increases to the future. For supertranslations, the is flux also non-negative if $s(x^A) >0$, i.e., $S^a$ is a `time-like supertranslation' that moves every cross-section $C$ of $\N$ to a cross-section $C^\prime$ that lies entirely to its future. One can interpret the  $2\F^{(2)}_{d}$ and $2\F^{(2)}_{s}$ as the `energy and super-momentum' carried by (weak) gravitational waves across the perturbed NEH (after taking into account the factor of $2$ associated with the Noether charge). \\
\goodbreak

\emph{Remarks:}

(i) The boundary conditions introduced in Sec.~\ref{s2} ensure that $\N$ becomes an NEH in the distant future, i.e. the expansion $\theta_{(\ell)}$ goes to zero as $v\to \infty$ for all $g \in \ps$. In binary coalescence or gravitational collapse, this is the NEH of the final remnant. At late times, it is natural to regard the space-time metric as a perturbation of the final Kerr black hole, and approximate the physical DH by a perturbation of the NEH of the Kerr remnant. If we keep terms to second order, the Raychaudhuri equation (\ref{Ray}) implies that the time derivative $\Lie_{\ell} \theta_{(\ell)}$ of the expansion of the perturbed NEH $\N$ is given by $- |\sigmaone_{mn}|^2\,$. Since $\Lie_{\ell} \theta_{(\ell)}$ is non-positive and vanishes in the remote future, $\theta_{(\ell)}$ is non-negative on $\N$. Consider any cross-section $C$ on which $\sigmaone_{mn}$ is non-zero, i.e., there is gravitational radiation. Then the area of any cross-section $C^\prime$ to its future satisfies $A [C'] > A [C]$. Thus, the area increases to the future, just as it does on a DH \cite{ak-dh}. This is just what one would expect from the fact that the perturbed $\N$ represents a weakly evolving DH. (With our conventions, as with DHs, the positive flux across $\N$ is inward pointing; the radiation is falling into the black hole.)

(ii) Since the flux $\F_d^{(2)} \ge 0$, the (second order truncated) dilation charge 
\be Q_d (\lambda)\, =\, Q^{(0)}_d +\lambda Q^{(1)}_d + \textstyle{\f{1}{2}} \lambda^2 Q^{(2)}_d\,  \ee 
increases with time. Consider a compact binary coalescence (or gravitational collapse) and let us start from the final remnant and move \emph{to the past} along $\N$. Then the area and the dilation charge decrease (to second order). However, when the decrease is non-negligible relative to the values associated with the final remnant, the second order truncation becomes inadequate and one can no longer regard the perturbed NEH as a reliable substitute for the full DH. Close to the merger, for example, one would have to abandon $\N$ as the internal boundary, and use the DH instead.

(iii) For the supermomentum charge, within the domain of validity of the second order perturbation theory we have that, for supertranslations $S^c = \so\ello^a$ with $\so\geq 0$
\be \label{supermomentum} Q_S (\lambda)\, =\, Q^{(0)}_S +\lambda Q^{(1)}_d + \textstyle{\f{1}{2}} \lambda^2 Q^{(2)}_S\, =\,  -\,{\f{\lambda^2}{16\pi G}}\, \oint_C  \thetatwo\,S^c\,\,\epsiloncirc_{cab}\, \le \, 0\, , \ee
where in the last step we have used the fact that $\theta^{\prime\prime}_{(\ell)}$ is non-negative on $\N$. Note that even when the perturbed shear is zero to the past of a cross section $C$, the perturbed expansion does not go to zero there; \emph{we do not have an unperturbed NEH in the past.} This may seem counter-intuitive at first since one might consider starting with an unperturbed NEH in the past and then sending in a perturbation. In that case, the expansion would be non-zero in the asymptotic future, whence that space-time would not be in the phase space $\ps$ we considered. For space-times in $\ps$ all supermomentum charges vanish in the asymptotic future and those with $s>0$ are negative on $\N$, given by (\ref{supermomentum}).\\

Finally let us consider the fluxes associated with rotations and boosts. A direct calculation would require equations of motion relating the time derivatives of perturbations of $\beta_A$ on $\N$ with expansion and shear. A more convenient approach is provided by the flux formula (\ref{flux}) in which the equations of motion are `built-in'. Using the pullback to $\N$  of $\ThetaH_{abc}$ one obtains that for any $g\in \ps$ the flux $\F_\xi[\N_{1,2}]$ for \emph{any} symmetry vector field $\xi^a \in \g$ is given by
\be \label{eq:flux_generic}\F_\xi[\N_{1,2}]\,=-\, \f{1}{16\pi G} \int_{\N_{1,2}} \, \Lie_{X}q^{mn}\left(\sigma^{(\ell)}_{mn} - \f{1}{2}\theta_{(\ell)} q_{mn}\right)\epsiloncirc_{abc}
\, \ee
(see, e.g., Eq.~6.26 of \cite{cfp}). Here, as before, indices are raised and lowered using the background \emph{2-metric} $\qcirc_{mn}$. Note that this expression is independent of the choice of $\ell^a$, and the choice of inverse of $\qcirc^{mn}$. 
Moreover it is manifest that the flux vanishes if $g\in \psneh$, as expected. To obtain the flux on the perturbed NEH, as before, we take a parameter family of solutions $g(\lambda)\in\ps$ passing through a $\gcirc \in \psneh$, differentiate with respect to $\lambda$ and evaluate the result at $\lambda=0$. As an independent check on results we already obtained, let us begin by examining the first order flux. Taking the first derivative with respect to $\lambda$, and evaluating at $\lambda=0$ using the fact that  ${}^\circ\theta_{(\ell)}\, \=\,0\, \= \,\thetaone$ we obtain 
\be \F^{(1)}_{\xi} [\N_{1,2}]\, =\, -\f{1}{16\pi G}\, \int_{\N_{1,2}} \Lie_{X}\qcirc^{mn}\left(\sigmaone_{mn}\right)\epsiloncirc_{abc}\, .\ee
Now, because $X$ is a conformal Killing vector of\, $\qcirc_{ab}$\, and\, $\qcirc^{mn}\sigmaone_{mn} = -q^{\prime\,mn}\, {}^\circ\sigma^{(\ell)}_{mn}\, \=\, 0$\,  on $\psneh$, this term vanishes. Therefore, to the first order the flux vanishes for \emph{all} NEH symmetries $\xi^a$, confirming our earlier result, now without reference to a chart and without having to explicitly invoke Einstein's equations (since they are already built-in at the start). 

Hence the leading order contribution to the flux comes at the second order. To the second order the only non-zero terms that contribute to the flux are 
\be \F^{(2)}_{\xi} [\N_{1,2}]\, =\, - \f{1}{16\pi G}\, \int_{\N_{1,2}} \left[\Lie_{X}\qcirc^{mn}\left(\sigma^{\prime\prime\,(\ell)}_{mn} - \f{1}{2}\thetatwo q_{mn}\right) + 2 (\Lie_X q^{mn})'\sigmaone_{mn}\right]\epsiloncirc_{abc}\, .\ee
All other terms vanish since ${}^\circ \sigma_{ab}^{(\ell)},\,{}^\circ\theta_{(\ell)}$ and $\thetaone$ vanish on $\N$. Using again the relations   $\Lie_X q^{mn} \,\= \, -2\phi q^{mn}$ and $q^{mn}\sigma^{\prime\prime\, (\ell)}_{mn} = -2q^{mn\,\prime}\sigmaone_{mn}$, we obtain
\begin{align}
\label{flux2}
\F^{(2)}_{\xi} [\N_{1,2}]\, &=\,- \f{1}{16\pi G}\, \int_{\N_{1,2}} \left[-2\phi
\left(2h^{mn}\sigma^{\prime\,(\ell)}_{mn} - \f{1}{2}\thetatwo q_{mn}\right) + 2 (\Lie_X h^{mn})^\prime\sigmaone_{mn}\right]\epsiloncirc_{abc}\, ,\\
&=\, \f{1}{16\pi G}\, \int_{\N_{1,2}} \Big[(\Lie_{\xi} \hone_{mn})\, (\dot\hone^{mn})\, + \phi \dot{(\ubhtwo)}\,\Big] \epsiloncirc_{abc}\,,
\end{align}
where we have used Eq.~\eqref{firstorder2} and Eq.~\eqref{secondorder}, and where indices are raised and lowered using the background \emph{2-metric} $\qcirc_{ab}$.
In particular, we see that for supertranslations and dilations we recover Eq.~\eqref{flux1}, previously obtained using the Raychaudhuri equation on source-free solutions to Einstein's equations. For rotations (\ref{flux2}) implies
\be \label{flux3} 
\F^{(2)}_{R} [\N_{1,2}]\, =\, \f{1}{16\pi G}\, \int_{\N_{1,2}} \Big[(\Lie_{R} \hone_{mn})\, (\dot\hone^{mn})\, - \, {\textstyle{\f{1}{2}}}\,(D_m R^m)\, \dot{(\ubhtwo)}\,\Big] \epsiloncirc_{abc}\, .\ee
The first term in\, $\F^{(2)}_{R}$\, has the same form as in the angular momentum flux formula at $\scrip$ for perturbations off a stationary background. However, unlike in the expression of the supermomentum flux, there is now an extra term, proportional to $D_aR^a$ which is generically non-zero because, while $R^a$ is a Killing field of the round 2-sphere metric $\qo_{ab}$, it need not be a Killing field of the physical metric $q_{ab}$. At $\scrip$, this term vanishes because in the parallel calculation $q_{ab} = \qo_{ab}$ there. Returning to perturbed NEHs, note that if $R^a$ were to be a Killing field of the physical metric $q_{ab}$, and if the perturbations were also to be axisymmetric, then the flux $\F^{(2)}_{R} [\N_{1,2}]$ vanishes \emph{even when the perturbation is time dependent}, just as one would expect. Similarly, even if $R^a$ is not a Killing field of $q_{ab}$, the flux vanishes if the perturbations are time independent, again as one would expect on physical grounds. Both these properties mirror what happens at $\scrip$.

For a boost vector field $B^a$ let us denote, as before, by $\t{B}^a$ the `horizontal part'\,\,  --i.e., the part that is tangential to the $v={\rm const}$ cross-sections of $\N$--\,\, and by $n_a$ the 1-form orthogonal to these cross-sections. Then the flux is given by 
\be \label{flux4}
\F^{(2)}_{B} [\N_{1,2}]\, =\, \f{1}{16\pi G}\, \int_{\N_{1,2}}\!\! \Big[  
(\Lie_{B}\, \hone_{mn})\, (\dot\hone^{mn})\, - \, {\textstyle{\f{1}{2}}}\,(D_m \t{B}^m)\, (\dot{\ubhtwo})\Big]\, \epsiloncirc_{abc}\, . \ee
Again, the flux vanishes if the perturbations are time independent, as one would expect physically.

\section{Discussion}
\label{s5}

While there is significant literature on NEHs, by and large the focus has been on the structure and symmetries of \emph{individual} NEHs. Just as generic space-times do  not admit any symmetries and, in 4-dimensions, the dimension of the isometry group of any given metric is less than 10, a generic NEH does not admit any symmetry and the symmetry group of any given NEH can not exceed 5 \cite{lp1}.  In the companion paper \cite{akkl1},  we shifted the focus from the structure and symmetries admitted by \emph{individual} NEHs to a structure \emph{shared by all} NEHs, and the group that preserves that `universal structure'. We found that the universal structure is closely related to --but slightly weaker- than that at null infinity, $\scri$, of asymptotically flat space-times. Consequently, the NEH symmetry group $\G$ turned out \emph{infinite dimensional}, a 1-dimensional extension of the BMS group $\B$. The extension consists of adding a dilation symmetry that, in a certain sense, extends the 4-dimensional translation subgroup of $\B$ to a 5-dimensional group. Examples of some of the physically most interesting NEHs --with zero as well as positive cosmological constant--  show that this extension is inevitable. 

In this paper we used a covariant phase space framework to compute charges and fluxes associated with generators $\xi^a$ of the symmetry group $\G$ on unperturbed as well as perturbed NEHs. In recent years the standard covariant phase space-framework of general relativity \cite{cw,abr,wz} has been extended to space-times admitting null boundaries $\N$ (see in particular \cite{cfp,cp2019,speziale1,ww2021,freidel1}). We worked in that general paradigm but with several modifications motivated by the fact that we are primarily interested in the late stages of gravitational collapse and black hole mergers. Most important among these is our focus on the sub-manifold $\psneh$ of the full phase space $\ps$, consisting of those solutions $\gcirc_{ab}\in \ps$ for which the null boundary $\N$ is an NEH, $\H$. However, we allowed first and second order perturbations that do not preserve the NEH character of the boundary. As a result we were able to obtain charges and fluxes both for the background solution $\gcirc_{ab}$ for which $\N$ is an NEH, as well as generic first and second order perturbations of $\gcirc_{ab}$ under which it is no longer an NEH. To obtain the charges $Q_{\xi} [C]$ associated with any symmetry vector field $\xi^a$ on 2-sphere cross-sections $C$ of $\N$, one needs to extend the symmetry vector fields $\xi^a$ from the boundary $\N$ to vector fields $X^a$ in a neighborhood of $\N$ in the space-time manifold $M$. In Appendix~\ref{a2} we discuss this extension and its important properties; in particular, the extensions preserve the symmetry algebra $\g$.

As one would expect on physical grounds, all the fluxes vanish \emph{for the background} $\gcirc_{ab}$ for which the boundary $\N$ is an NEH $\H$, whence the charge integrals are independent of the 2-sphere cross section of $\H$ on which they are evaluated. The background charge $Q_d^{(0)} [C]$ associated with the dilation vector $d^a$ on $\H$ is positive  and, in the case when $d^a$ is the restriction to $\H$ of a Killing vector, it equals the Komar integral (modulo a fixed multiplicative constant). The same is true for a rotation vector field $R^a$ in $\g$. The charges associated with all supertranslations vanish. As one would expect, non-trivial fluxes arise when we keep terms that are second order in perturbations. To this order, generically all charges and fluxes are non-zero. The flux associated with the dilation vector field is non-negative and vanishes only if the perturbation is time-independent --just as one would hope, since it can be interpreted as  ``energy carried by perturbations across $\H$". Fluxes associated with supertranslations $S^a = \so(\theta,\varphi)\,\ello^a$ are also positive if $\so(\theta, \varphi)$ is positive. Charges and fluxes associated with vector fields generating Lorentz transformations in $\G$ also have physically expected properties. For example, fluxes vanish for the background as well as to first order in perturbations, and they vanish also to second order if the linearized shear vanishes.

These physically expected properties hold because the boundary $\N$ is required to be an NEH or a perturbed NEH. On the full phase space $\ps$ considered in \cite{cfp}, for example, the positivity properties just mentioned fail. Also, in that framework one can take $(M, g_{ab})$ to be the exterior of the null cone of a point in \emph{Minkowski space} (or a part thereof), with the null cone (or part thereof) serving as the boundary $\N$. Then the charge and flux associated with the dilation vector field (which is also a symmetry in the Lie-algebra of \cite{cfp}) are non-zero. Given that space-time under consideration is \emph{flat}, it is difficult to associate physical significance to these charges and fluxes, and similarly, at generic points the phase space $\ps$ that requires the boundary only to be null. These awkward situations do not arise on $\psneh$. Thus, the NEH framework carves out a sector of the full phase space framework where formulas for charges and fluxes are physically meaningful. 

Ref. \cite{akkl1} also extended the known results on NEHs in another direction by introducing the notion of multipole moments, without having to assume axisymmetry as in previous discussions (see, e.g., \cite{aepv,Ashtekar_2013}). These moments are a set of numbers --calculated without using any extraneous structures such as coordinates, frames or gauge choices-- that provide an invariant characterization of the intrinsic geometry of the horizon, and its angular momentum structure. By now there have been multiple studies \cite{Gupta_2018,pookkolb2020horizons,prasad2021tidal,Mourier:2020mwa} of numerical simulations of binary black hole coalescences which show, through a study multipole moments, that soon after the common DH is formed, it can be well approximated by a perturbed NEH. The rate of change of multipole moments provides an invariant description of horizon dynamics both for DHs and perturbed NEHs. Using the close similarity between structures on perturbed NEHs and those at null infinity, a forthcoming work will show that this horizon dynamics can be deduced directly from the the post-merger waveforms at null infinity, $\scrip$. The idea that the horizon dynamics in the strong field region could perhaps be read-off from waveforms in the weak field region has been put forward \cite{akrev} and pursued in some detail in the literature \cite{schnetter_2006,Gupta_2018,pookkolb2020horizons,prasad2021tidal}. The goal of a  forthcoming work is to provide concrete steps to develop this \emph{gravitational wave tomography} in a systematic fashion.

However, using perturbed NEHs one cannot hope to describe the regime in which fully non-linear, dynamical effects are important, e.g., during or immediately after a black hole merger. As discussed in \cite{akkl1}, the `obvious' strategy of replacing perturbed NEHs with event horizons is unlikely to work because, as the Vaidya metric illustrates, event horizons can also grow in regions where the space-time metric is flat \cite{akrev}. Therefore, charges associated with event horizons will also have spooky features. In our view, to describe the strongly non-linear phase close to the merger itself, it would be more fruitful to consider the covariant phase space of solutions of Einstein's equations that admit a DH as the inner boundary. Thus, the boundary would no longer be null; it would be space-like in the strongly non-linear phase, but can be approximated by a null NEH soon after the merger. While construction of the corresponding phase space framework would be non-trivial, thanks to the rich set of results on DHs, it is within reach.
 
In quantum considerations, on the other hand, the exact DH describing black hole evaporation would be time-like and its area would therefore shrink in time \cite{ak-dh,aa-eva}. Nonetheless, it would be approximated extremely well by a perturbed NEH for a very long time, e.g., over the $\sim 10^{76}$ years it takes a solar mass black hole to shrink to a lunar mass through Hawking evaporation. It is only when the black hole enters a genuinely quantum gravity regime that this description would fail; but so would any description based on classical general relativity! The perturbed NEH framework developed in this paper is likely to be useful in the very long phase during which semi-classical approximation holds.  The multipole moments of  Sec.~\ref{s2} of \cite{akkl1} --particularly, their slow evolution due to perturbations-- as well as the close similarity between the NEH symmetry group $\G$ and the BMS group $\B$  are likely to be useful in analyzing (at least) this phase of the evaporation process. This approach is in the same spirit as the one in \cite{hps} that was developed independently, albeit some important technical differences.

\section*{Acknowledgment}
This work was supported by the NSF grant PHY-1806356, the Eberly Chair funds of Penn State, and the Mebus Fellowship to NK. AA thanks Simone Speziale and Wolfgang Wieland for discussions on the covariant phase space of general relativity in presence of null boundary in 2017. MK was financed from budgetary funds for science in 2018-2022 as a research project under the program "Diamentowy Grant". MK and JL were supported by Project OPUS 2017/27/B/ST2/02806 of Polish National Science
Centre. 

\begin{appendix}

\section{Charges and Fluxes: The Conceptual Framework}
\label{a1}

For convenience of the reader, in this Appendix we begin by briefly recalling the relevant features of the covariant phase space framework for general relativity \cite{cw,abr,wz}. This summary will provide a broader perspective for our results on the charges and fluxes on perturbed NEHs, and also for the more recent discussions of the covariant phase space in presence of internal null boundaries \cite{cfp,cp2019,speziale1,ww2021,freidel1}.

Consider first the covariant phase space of globally hyperbolic solutions of Einstein's equations without internal boundaries (and satisfying the standard asymptotic fall-off conditions). It admits a pre-symplectic structure $\mathbf{\Omega}$ which can be derived systematically starting from the Lagrangian. Thus, given any vacuum solution $(M, g)$ of Einstein's equations, and two linearized solutions $\delta_1 g$ and $\delta_2 g$ on this background,\,  $\mathbf{\Omega}|_{g}(\delta_1 g,\, \delta_2g)$\, is the action of the pre-symplectic structure on the tangent vectors $\delta_1 g,\, \delta_2g$ at the point $g$ of the covariant phase space. Since $\mathbf{\Omega}$ is an exact 2-form, it admits 1-form potentials $\mathbf{\Theta}$;\, $\mathbf{\Theta}|_{g} (\delta g)$ is the action of the potential on a tangent vector $\delta g$ at $g$. Thus, on the phase space we have: $\mathbf{\Omega}(\delta_1 g,\, \delta_2g) = \Lie_{\delta_1 g} \, (\mathbf{\Theta}(\delta_2 g))\, -\, \Lie_{\delta_2 g}\, (\mathbf{\Theta}(\delta_1 g))\, -\, i_{[\delta_1 g,\, \delta_2 g]}\,\mathbf{\Theta}$, where $i$ denotes the contraction of a vector field with a 1-form on the phase space, and the bracket denotes the Lie bracket of two tangent vector fields thereon. Our primary interest lies in the Hamiltonians and charges that correspond to symmetry vector fields $X^a$ in space-time, evaluated at specific points $g$ in the phase space. We will follow the practice adopted in the literature and work in a neighborhood of the point $g$ of interest and consider vector fields $\delta g$ in that neighborhood which: (i) commute with a given vector field $\delta_X\, g \equiv \Lie_{X}\, g_{ab}$ on the phase space, and, (ii) span the tangent space at each point in that neighborhood. In this case, one has\, $\mathbf{\Omega}(\delta g,\, \delta_X g) = \Lie_{\delta g} \, (\mathbf{\Theta}(\delta_X\, g))\, -\, \Lie_{(\delta_X\, g)}\, (\mathbf{\Theta}(\delta g))\, $. 
Now, if the 1-form $i_{(\delta_X g)}\,\mathbf{\Omega}$ is exact, i.e., if\, $\mathbf{\Omega}(\delta g,\, \delta_X g) = \Lie_{\delta g}\, H_X(g)$ for some function $H_X(g)$ on the phase space, then (modulo an additive constant) $H_X(g)$ would be the Hamiltonian generating the canonical transformation defined by the symmetry vector field $X^a$. 
For example, if one considers solutions $(M, g_{ab})$ that are asymptotically flat at spatial infinity ( and with no internal boundaries), then the Hamiltonians $H_X(g)$ are 2-sphere integrals at spatial infinity, representing conserved charges $Q_\xi (g)$ associated with asymptotic symmetries \cite{cw,abr}.  

The situation is more involved in our case since space-times in $\ps$ have an \emph{internal boundary} $\N$: in presence of this boundary, the analog of the 1-form $i_{(\delta_X g)}\,\mathbf{\Omega}$ generically fails to be exact on $\ps$ (unless $X^a$ vanishes on the boundary). Nonetheless, as shown in \cite{cfp,cp2019,speziale1,ww2021,freidel1}, one can systematically associate charges $Q_\xi [C]$ with each symmetry vector field $\xi^a$ and a cross-section $C$ of $\N$. We will now summarize this procedure.

Note first that the action of the pre-symplectic 2-form $\mathbf\Omega$ on tangent vectors $\delta_1 g$ and $\delta_2 g$ at the point $g$ in the phase space is given by the integral over any Cauchy surface of a closed 3-form  $\j_{abc} (g; \delta_1 g, \delta_2 g)$ \emph{in space-time,} that depends on $g$ and $\delta_1 g,\, \delta_2 g$. The Lagrangian formulation of general relativity provides us with the explicit expression of this \emph{symplectic current}, which is spelled out in Appendix~\ref{a2}. The pre-symplectic potentials $\mathbf\Theta$ are also expressible as integrals of 3-forms $\Theta_{abc} (g; \delta g)$ over a Cauchy surface. Thus, while $\mathbf\Theta$ is a 1-form \emph{in the phase space}, $\Theta_{abc} (g; \delta g)$ is a 3-form \emph{in space-time} (that depends on the given solution $g_{ab}$, and a linearized solution $\delta g_{ab}$ on that background). To arrive at the expression of charges, one works directly with the 3-forms $\j_{abc} (g; \delta_1 g, \delta_2 g)$ and  $\Theta_{abc} (g; \delta g)$, rather than the pre-symplectic structure $\mathbf\Omega$ and its potentials $\mathbf{\Theta}$. Secondly, although charges are associated with symmetry vector fields $\xi^a$ defined just on $\N$, the procedure requires us to extend them to vector fields $X^a$ in a neighborhood of $\N$ in the space-time manifold $M$. This issue is discussed in detail in Appendix~\ref{a2}.%
\footnote{One needs this extension only to a neighborhood of the internal boundary. In calculations of charges and fluxes on $\N$, it is convenient to work with extensions that vanish away from this neighborhood. In terms of Fig.~\ref{fig:triangle}, this strategy enables one to ignore the boundary of partial Cauchy surfaces $\Sigma$ at infinity and focus just on their interior boundaries $\partial \Sigma$.}

Consider now a partial Cauchy surface $\Sigma$ that intersects $\N$ in a cross-section $C = \partial\Sigma$ (as in Fig.~\ref{fig:triangle} where $\N$ is assumed to be an NEH $\H$). Although $\int_\Sigma\, \j_{abc} (g;\, \delta g, \delta_X g)$ is generically not of the form\, $\Lie_{\delta g} H_X (g)$,\, one can add to it  a \emph{surface term}\, $\oint_{C} X^a\, \Theta_{abc}(g;\, \delta g)$\, so that $\int_\Sigma \j_{abc} (g;\, \delta g, \delta_X g)\, + \, \oint_{C} X^a \Theta_{abc}(g;\, \delta g)$\, is the exact variation of a phase space function $ Q_X(g)$,
\be \label{QX}
\int_\Sigma \j_{abc} (g;\, \delta g, \delta_X g)\, + \, \oint_{\partial\Sigma} X^a \Theta_{abc}(g;\, \delta g)\, =\, \Lie_{\delta g} Q_X (g)\,[C] \equiv \,\delta Q_X(g)\, [C] \, ,
\ee
provided $\Theta_{abc} (g;\, \delta g)$ is chosen covariantly (see Eq. (\ref{covariance})). As the notation  $ Q_X(g) [C]$ suggests, $ Q_X(g)$ does not depend on what $\Sigma$ does away from $\N$, but depends only on fields evaluated on the cross-section $C$ at which $\Sigma$ intersects $\N$. It is the charge associated with the symmetry $X^a$, evaluated at $C$. The charge does not generate 
a canonical transformation that implements the symmetry on $\ps$ but there is a proposal \cite{Barnich:2011mi, Wieland:2021eth} to modify the phase space structure to improve on this situation.
\smallskip

Thus, as mentioned in Sec.~\ref{s2.2}, one requires two inputs to arrive at the expression (\ref{QX}) of charges.  First, one needs to choose a certain extension $X^a$ of our symmetry vector fields $\xi^a$ on $\N$ to the space-time interior, that is discussed in Appendix~\ref{a2}. The second input is a pre-symplectic potential $\mathbf{\Theta} (g;\,\delta g)$ that appears in the second term on the left side of (\ref{QX}) (as well as in the expression of $Q_X$ that appears on the right side, which is spelled out in Sec.~\ref{s2.2}). As on any phase space, these potentials are not unique; one can add to $\Theta_{abc}(g; \delta g)$ a term of the type $\delta f_{abc}(g)$. The procedure requires us to uniquely single out a preferred 3-form $\ThetaH_{abc}(g; \delta g)$ \emph{on the boundary $\N$} by eliminating this freedom. The required choice was provided in \cite{morales} in a somewhat different context, and obtained independently in \cite{cfp} by a generalization of the set of conditions laid out in \cite{wz} to accommodate the fact that $\N$ is an internal null boundary rather than $\scrip$. On our $\ps$, the key conditions that select a canonical $\ThetaH_{abc}(g; \delta g)$ are the following:\smallskip

\noindent
(i) \emph{Covariance, locality and analyticity:} $\ThetaH_{abc}$ should depend only locally and analytically on $g_{ab}, \delta g_{ab}$ and a finite number of their derivatives, and fields made available by our boundary conditions defining $\ps$. (Thus, it cannot depend on any non-dynamical structures introduced by hand, such as coordinates, tetrads or a choice of an $\ell^a$ in the equivalence class $[\ell^a]$). In particular, given a vector field $X^a$ on $M$ that is tangential to the boundary $\N$, the 3-form $\ThetaH_{abc}$ on $\N$ must satisfy: 
\be \label{covariance}
\Lie_{(\delta_X g)}\, \ThetaH_{abc}(g;\,\delta  g) \equiv  \delta_{X}\, \ThetaH_{abc}(g;\,\delta  g)
\, =\, \Lie_{X}\,\big( \ThetaH_{abc}(g;\,\delta  g)\big)\, ,\ee
where, as before, $X^a$ is an appropriate extension to space-time of the symmetry vector field $\xi^a$ on $\N$. 

\noindent
(ii) \emph{Time independence on an NEH:} If $\N$ is an NEH for a given $g$, then $ \Lie_\ell\, \ThetaH_{abc}(g; \delta g) = 0$ for arbitrary tangent vectors $\delta g$ at that $g$. This condition is motivated by the fact that freely specifiable fields that determine the NEH geometry -and physics on NEHs- is time independent. \smallskip

This preferred 3-form $\ThetaH_{abc}\,(g; \delta g)$ on $\N$ is constructed by the following two step procedure. First, one notes that the `obvious' symplectic potential $\Theta_{abc}(g;\delta_1 g, \delta_2 g)$ in the bulk, obtained applying the variational principle to the Einstein-Hilbert Lagrangian density, is:  
 \be  \Theta_{abc}(g;\delta g) = \frac{1}{16\pi G}\,\,\epsilon_{abc}{}^d\,(g^{ef}\nabla_d\,\delta g_{ef}-\nabla^e\delta g_{ed})\, .  \ee
Denote its pull-back to $\N$ by $\Theta_{\pb{abc}} (g, \delta g)$. As noted in Sec.~\ref{s2.2}, the desired potential $\ThetaH_{abc} (g, \delta g)$ can then be expressed as \cite{morales,cfp} 
\be \label{ThetaH} 
\ThetaH{}_{abc} (g; \delta g) := \, {\Theta}{}_{\pb{abc}} (g; \delta g) - \f{1}{8\pi G}\, 
\delta \big( (\theta_{(\lcirc)}\, \epsilon^\circ_{abc})(g)\big)\, . \ee 
As in Sec.~\ref{s2.2}, $\theta_{(\ell)}$ denotes the expansion of any geodesic null normal $\ell^a$ to $\N$,\, and $\epsilon_{abc}$, the intrinsic volume 3-form on $\N$ defined in terms of the space-time volume 4-form $\epsilon_{abcd}$ via 
\be \epsilon_{abc} := n^d \epsilon_{dabc} \quad \hbox{\rm where $n^a$ is any vector field on $\N$ s.t. $\lcirc^a n^b\, g_{ab} =-1$} \, .\ee

\section{Extension of symmetry vector fields on the boundary $\H$ to the bulk}
\label{a2}

As noted in Sec.~\ref{s2.2}, to calculate the Noether charge, one needs the `radial' derivative of the symmetry vector field, requiring us to extend the symmetry vector fields $\xi^a$ on the null boundary to the interior to leading order in $r$. In this Appendix we will provide the expression of this extension $X^a$ in Newman-Unti type chart introduced in Sec.~\ref{s2.1.1}, and discuss its most useful properties.

Let us begin with the obvious geometric conditions on $X^a$ stemming from the fact that, being an extension of $\xi^a$ on $\H$, $X^a$ has to be tangential to $\H$. Since $\H$ is given by $r=0$ in our chart, in a neighborhood of $\H$,\, $X^a$ must have the form
\begin{equation}
 \label{X2}   X = X^v \partial_v + H^A\partial_A + r \tilde{X}^r \partial_r,
\end{equation}
where $X^v, H^A, \tilde{X}^r$ are all smooth functions of $v,r,x^A$. ($H^a$ is `horizontal' in the sense that it is tangential to the constant $v$ and $r$ 2-spheres.) Since the restriction of $X$ to $\H$ is, by definition, an NEH symmetry $\xi^a$, from (\ref{xi2}) we have
\be \label{X3} X^v\, \=\,\, (k+\phi(x^A))\, v + s(x^A),\qquad H^A{}_{,v} \,\, \= \,0, \qquad {\rm and} \qquad \Lie_{H} q_{ab}\,\, \=\,\, 2\,\phi(x^A)\,\, q_{ab}\, .  \ee
In particular, $H :\=\, H^A\partial_A$ is a conformal Killing field of $q_{AB}$ on the NEH. 

As Eq.~(\ref{QN2}) shows, to define the Noether charge we also need the leading part of $\tilde{X}^r|_{\H}$, in addition to $X^a|_\H = \xi^a$. We will now show that this  leading term is determined by the following condition that is satisfied by all $g\in \ps$: Given \emph{any} null geodesic normal $\lcirc^a$, the restriction to $\H$ of the 1-form $g_{ab}\lcirc^a$ obtained by lowering its index by the metric is independent of $g \in \ps$ (see Sec.~\ref{s2.1.1}). The desired extension $X^a$ generates a 1-parameter family of diffeomorphisms on the space-time manifold $M$ that leaves $\H$ invariant. The extension must be such that the 1-parameter family of metrics $g_{ab}(\lambda)$ and geodesic null normals $\lcirc^a(\lambda)$, obtained starting from any fiducial $g^{\circ} \in \psneh$ must be in $\psneh$ and satisfy $g_{ab}(\lambda) \lcirc^b(\lambda)\, \=\, g^{\circ}_{ab} \lcirc^b(\lambda)$. Therefore, the extension must satisfy%
\be \big(\Lie_X g_{ab}\big)\, \lcirc^b\, \=\, 0,\quad \hbox{\rm and hence}\quad 
\Lie_X l_{a}\, \=\, -(k +\phi)\, l_{a}\, . \ee
Using the form (\ref{metric}) of the metric in a neighborhood of $\N$ it is easy to verify that this condition determines $\t{X}^r$ to leading order:
\be \t{X}^r\, \=\, - (\phi +k) \quad \hbox{\rm that is,}\quad \t{X}^r = - (\phi +k) + O(r) \ee
Thus, in our chart, the extension must have the form:
\be \label{Xfinal}
    X = \big((k+\phi)v + s\big) \partial_v + H^A \partial_A \, +\, r\, \big(-(k +\phi)\,\partial_r + \t{X}^v\, \partial_v + \t{X}^A\, \partial_A\big)\, +\, r^2\, \t{\t{X}}^r\, \partial_r\,. \ee
Here $\phi$ and $s$ are functions only of angular coordinates $x^A$ defined by the NEH symmetry vector field $\xi^a$, and $\t{X}^v\,\t{X}^A\,$ and $\t{\t{X}}^r\,$ are arbitrary, smooth functions of all 4 coordinates. As is clear from our discussion in Secs.~\ref{s3} and \ref{s4}, the charges and fluxes are insensitive to the choice of these undetermined functions.

We will conclude by noting three interesting properties of this extension.

(i) If a $g \in \ps$ admits a Killing vector $K^a$ that is tangential to the horizon, then its restriction to $\H$ is of course an NEH symmetry vector field $\xi^a$. The extension $X^a$ of $\xi^a$ is such that $K^a$ agrees with $X^a$ to $O(r)$. Thus, a desired interplay between the form of $X^a$ and the space-time symmetries holds. This result implies that $\H$ happens to be a Killing horizon for $K^a$ with surface gravity $k$, then  $X := k (v\partial_v - r \partial_r)\, + O(r^2)$.

(ii) The vector fields $\delta_X g$ on $\psneh$ also have a desirable property that involves the symplectic current. In this discussion we will only need first order perturbations. Therefore, to simplify notation, let us set $\delta g_{ab} = h_{ab}$ (rather than $\hone_{ab}$ as in Sec.~\ref{s4}). Given a solution $g_{ab}$ to the vacuum Einstein's equations (possibly with a cosmological constant) and two linearized solutions $h_{ab}$ and $h^\prime_{ab}$ that satisfy the linearized Einstein equations, the Einstein-Hilbert Lagrangian leads us to a symplectic current vector field $\j^a (g;\, h,h^\prime)$ in space-time \cite{cw,abr,wz}:
\begin{equation}
    \j^a(g;\,h, h^\prime) = \frac{1}{16 \pi G}\,  P^{abcdef}\, \left(
    h^\prime_{bc} \nabla_d h_{ef} - h_{bc} \nabla_d h^\prime_{ef}
    \right),\label{sympl}
\end{equation}
where
\begin{equation}
    P^{abcdef} = g^{ae} g^{bf} g^{cd} - \frac{1}{2} g^{ad} g^{be} g^{cf} - \frac{1}{2} g^{ab} g^{cd}g^{ef} - \frac{1}{2} g^{bc} g^{ae} g^{df} + \frac{1}{2} g^{bc} g^{ad} g^{ef}\, . 
\end{equation}
The symplectic 3-form $\j_{abc}$ of  Appendix~\ref{a1} is the dual of this current: $\j_{abc} = \j^d \epsilon_{dabc}$. The second property of the extension is that, for all $\g \in \psneh$, and time independent perturbations $h$, the pull-back to $\N$ of the symplectic 3-form vanishes identically
\be \j_{\pb{abc}} (g;\, h, \Lie_X g)\, \= \, 0 \, .\ee
Consequently, the symplectic flux $\int_{\cal P} \j_{abc} (g;\, h, \Lie_X g)$ vanishes across \emph{any patch} $\cal{P}$ of $\N$. 

(iii) The third property has to do with the Lie algebra structure of the extended symmetries. In view of the fact that charges and fluxes depend only on the `values of $X^a$ and their first transversal derivatives' at $\N$, we are led to introduce an equivalence relation on these vector fields: $X_1 \approx X_2$ if and only if: \\
\centerline{ $X_1\, \= \, X_2 \qquad {\rm and}\qquad X_1^r\, \=\, X_2^r\quad$ in Eq.~(\ref{X2})}
Let us denote by $\{X\}$ the equivalence class to which $X$ belongs. Then it is easy to check that the Lie-bracket on the space of these equivalence classes induced by the commutator is well-defined:
\be \label{comm} 
[\{X_1\},\, \{X_2\}] \, :=\, \{[X_1,\, X_2]\} \qquad \hbox{\rm for any choice of $X_1 \in \{X_1\}$ and $X_2 \in \{X_2\}$}\, . \ee 
Furthermore, this Lie algebra of equivalence classes $\{X\}$ of extensions is isomorphic with the Lie algebra $\g$ of symmetry vector fields $\xi^a$.

To summarize, conditions used to define the phase space $\ps$ imply that there is a class of natural extensions of the NEH symmetry vector fields $\xi^a$ to a neighborhood of $\H$. The charges and fluxes depend only on the information in $X^a$ that is completely contained in the vector field $\xi^a$ on $\H$. This class of extensions has three physically desirable properties; the first related to Killing fields that may be present in a neighborhood of $\H$, the second related to the symplectic 3-form, and the last related to the Lie algebra structure of the extended $X^a$. Therefore, the explicit form (\ref{Xfinal}) of these vector fields may be useful also in other contexts, in addition to the discussion of charges and fluxes.

\end{appendix}
\bibliographystyle{JHEP}
\bibliography{bibl2}
\end{document}